\documentclass[longauth]{aa} 
\usepackage{lscape}
\usepackage{natbib,twoopt}
\usepackage{graphicx}
\usepackage[usenames,dvipsnames]{xcolor}
\usepackage{comment}
\usepackage{bm}

\newcommand{\simgt}{\lower.5ex\hbox{$\; \buildrel > \over \sim \;$}}
\newcommand{\simlt}{\lower.5ex\hbox{$\; \buildrel < \over \sim \;$}}

\usepackage{txfonts}

\begin{document}

   \title{The eROSITA Final Equatorial-Depth Survey (eFEDS)}
    \subtitle{X-ray properties of Subaru's optically selected clusters}

   \author{N. Ota\inst{1, 2}
          \and
          N. T. Nguyen-Dang \inst{3}
          \and
          I. Mitsuishi\inst{4}
          \and
          M. Oguri\inst{5,6,7}
          \and
          M. Klein\inst{8}
          \and
          N. Okabe\inst{9,10,11}
          \and
          M. E. Ramos-Ceja\inst{12}
          \and
          T. H. Reiprich\inst{1}
          \and
          F. Pacaud\inst{1}
          \and
          E. Bulbul\inst{12}
          \and
          M. Br\"uggen\inst{13}
          \and
          A. Liu\inst{12}
          \and
          K. Migkas\inst{1}
          \and
          I. Chiu\inst{14,15,16}
          \and 
          V. Ghirardini\inst{12}
          \and
          S. Grandis\inst{8}
          \and
          Y.-T. Lin\inst{16}
          \and
          H. Miyatake\inst{7,17}
          \and
          S. Miyazaki\inst{18,19}
          \and
          J. S. Sanders\inst{12}
          }

   \institute{Argelander-Institut f\"{u}r Astronomie (AIfA), Universit\"{a}t Bonn, Auf dem H\"{u}gel 71, 53121 Bonn, Germany\\
              \email{naomi@astro.uni-bonn.de}
         \and
            Department of Physics, Nara Women's University, Kitauoyanishi-machi, Nara, 630-8506, Japan
         \and 
            Institut f\"{u}r Astronomie und Astrophysik T\"{u}bingen (IAAT), Universit\"{a}t T\"{u}bingen, Sand 1, 72076 T\"{u}bingen, Germany
        \and
            Graduate School of Science, Division of Particle and Astrophysical Science, Nagoya University, Furo-cho, Chikusa-ku, Nagoya, Aichi, 464-8602, Japan
        \and 
        Center for Frontier Science, Chiba University, 1-33 Yayoi-cho, Inage-ku, Chiba 263-8522, Japan
        \and 
        Department of Physics, Graduate School of Science, Chiba University, 1-33 Yayoi-Cho, Inage-Ku, Chiba 263-8522, Japan
       \and 
             Kavli Institute for the Physics and Mathematics of the Universe (Kavli IPMU, WPI), University of Tokyo, Chiba 277-8582, Japan
        \and
           Faculty of Physics, Ludwig-Maximilians-Universit{\"a}t, Scheinerstr. 1, 81679, Munich, Germany
        \and
            Physics Program, Graduate School of Advanced Science and Engineering, Hiroshima University, 1-3-1 Kagamiyama, Higashi-Hiroshima, Hiroshima 739-8526, Japan
        \and
            Hiroshima Astrophysical Science Center, Hiroshima University, 1-3-1 Kagamiyama, Higashi-Hiroshima, Hiroshima 739-8526, Japan
        \and
            Core Research for Energetic Universe, Hiroshima University, 1-3-1, Kagamiyama, Higashi-Hiroshima, Hiroshima 739-8526, Japan 
        \and         
            Max Planck Institute for Extraterrestrial Physics, Giessenbachstrasse 1, 85748 Garching, Germany 
        \and
            University of Hamburg, Hamburger Sternwarte, Gojenbergsweg 112, 21029 Hamburg, Germany
        \and
            Tsung-Dao Lee Institute, and Key Laboratory for Particle
Physics, Astrophysics and Cosmology, Ministry of Education,
Shanghai Jiao Tong University, Shanghai 200240, China
\and
Department of Astronomy, School of Physics and Astronomy,
and Shanghai Key Laboratory for Particle Physics and Cosmology,
Shanghai Jiao Tong University, Shanghai 200240, China
\and
Academia Sinica Institute of Astronomy and Astrophysics (ASIAA), 11F of AS/NTU Astronomy-Mathematics Building, No.1, Sec. 4, Roosevelt Rd, Taipei10617, Taiwan
\and
  Kobayashi-Maskawa Institute for the Origin of Particles and the Universe (KMI), Nagoya University, Nagoya, 464-8602, Japan
\and
National Astronomical Observatory of Japan, 2-21-1 Osawa, Mitaka, Tokyo 181-8588, Japan
\and
SOKENDAI (The Graduate University for Advanced Studies), Mitaka, Tokyo, 181-8588, Japan
             }

   \date{Received September 15, 1996; accepted March 16, 1997}

 
  \abstract
   {We present the results of a systematic X-ray analysis of optically rich galaxy clusters detected by the Subaru Hyper Suprime-Cam (HSC) survey in the eROSITA Final Equatorial-Depth Survey (eFEDS) field.
   }
   {Through a joint analysis of the SRG (Spectrum Roentgen Gamma)/eROSITA and Subaru/HSC surveys, we aim to investigate the dynamical status of the optically selected clusters and to derive the cluster scaling relations. }  
   {The sample consists of 43 optically selected galaxy clusters with a richness $>40$ in the redshift range of 0.16--0.89. We systematically analyzed the X-ray images and emission spectra using the eROSITA data. 
    We identified the brightest cluster galaxy (BCG) using the optical and far-infrared databases. We evaluated the cluster’s dynamical status by measuring three quantities: offset between the X-ray peak and BCG position, the gas concentration parameter, and the number of galaxy-density peaks.
   We investigated the luminosity-temperature and mass-luminosity relations based on eROSITA X-ray spectra and HSC weak-lensing data analyses. }
   {Based on these three measurements, we estimated the fraction of relaxed clusters to be $2(<39)$\%, which is smaller than that of the X-ray-selected cluster samples. After correcting for a selection bias due to the richness cut, we obtained a shallow $L-T$ slope of $2.1\pm0.5$, which is consistent with the predictions of the self-similar model and the baseline model incorporating a mass-concentration relation. The $L-M$ slope of $1.5\pm0.3$ is in agreement with the above-cited theoretical models as well as the data on the shear-selected clusters in the eFEDs field.} 
   {Our analysis of high-richness optical clusters yields a small fraction of relaxed clusters and a shallow slope for the luminosity-temperature relation. This suggests that the average X-ray properties of the optical clusters are likely to be different from those observed in the X-ray samples. Thus, the joint eROSITA and HSC observations are a powerful tool in extending the analysis to a larger sample and understanding the selection effect with a view to establish cluster scaling relations.}


   \keywords{Galaxies: clusters: intracluster medium; intergalactic medium; X-rays: galaxies: clusters}

   \maketitle
%
\section{Introduction}
One of the most powerful constraints on current cosmological models comes from observations of how the galaxy cluster population evolves over time. The redshift evolution of the cluster mass function, particularly in the number of high-mass clusters, is sensitive to cosmological parameters \citep{Allen11}. This makes observations of massive clusters in the distant universe vital. In particular, understanding the mass-observable scaling relations of these clusters is crucial to facilitating their use in cosmology \citep{Giodini13}. However, measurements at higher redshifts pose a considerable challenge due to the small sample size and difficulties in accounting for selection bias \citep{Pratt19}, along with the overall need for deep observations.

A combination of eROSITA and HSC observations has given us the unique opportunity to conduct a systematic study of optically selected clusters. The extended ROentgen Survey with an Imaging Telescope Array (eROSITA) on board the Spectrum-Roentgen-Gamma (SRG) satellite performed scanning X-ray observations of the $140~{\rm deg^2}$ eROSITA Final Equatorial Depth Survey field (the eFEDS field), during the performance verification phase \citep{Predehl21}. The eFEDS survey has improved X-ray data uniformity of optically selected clusters; the vignetted, average exposure time is $1.3$~ks in the 0.5--2~keV band \citep{Brunner22}. On the other hand, the Hyper Suprime-Cam \citep[HSC; ][]{Miyazaki18} Subaru Strategic Program (SSP) is a wide-field optical imaging survey \citep{Aihara18a,Aihara18b,Aihara22,Tanaka18,Bosch18}, which has a significant overlap with the eFEDS survey. From the HSC-SSP dataset, \citet{Oguri18} constructed a cluster catalog using the red-sequence cluster finding algorithm, CAMIRA \citep{Oguri14}. The catalog contains about 20,000 clusters at $0.1<z<1.3$ with richness $N>15$. The cluster mass-richness relation is carefully calibrated using weak-lensing analysis \citep{Okabe19,Murata19}. Therefore, by a joint analysis of the two surveys, the scaling relations and their dependence on cluster dynamical status can be studied over a wide range of redshifts and masses.

Recent X-ray follow-up observations of optically selected clusters have reported that the X-ray properties of the optical clusters are marginally different from those observed in the X-ray samples. \cite{Willis21} compared the XXL and CAMIRA catalogs, finding that 71/150 XXL clusters (67/270 CAMIRA clusters) are matched to the location of a CAMIRA cluster (an XXL cluster). Of the unmatched CAMIRA clusters, the stacked XMM-Newton data yielded a low, flat surface-brightness distribution, which is unlikely to follow the conventional $\beta$-model \citep{Cavaliere76}. From the XMM-Newton data analysis of 37 CAMIRA clusters, \cite{Ota20} found a small fraction of relaxed clusters compared to X-ray cluster samples suggesting that the optical cluster sample covers a larger range of the cluster morphologies. They also derived the luminosity-temperature relation and found that the slope is marginally shallower than those of X-ray-selected samples and consistent with the self-similar model prediction of 2 \citep{Kaiser86}. To obtain more conclusive results, we aim to improve the measurement accuracy and sample uniformity. 

In this paper, we study the dynamical status and scaling relations for a subsample of high-richness, optically selected clusters in the eFEDS field. This paper is structured as follows. Sectsions~\ref{sec:sample} and \ref{sec:optical} present the sample selection and optical analyses of the brightest cluster galaxies and weak-lensing mass, respectively. Section~\ref{sec:xray} describes the X-ray measurements of the cluster centroid and spectral properties. Section~\ref{sec:results} derives the centroid offset and the relations between mass and X-ray observables. Section~\ref{sec:discussion} presents a discussion of the implication of our results. 

The cosmological parameters used throughout this paper are $\Omega_{\rm m}=0.3$, $\Omega_\Lambda=0.7$ and $h=0.7$. We use the abundance table from \citet{Asplund09} in the X-ray spectral modeling. The quoted errors represent the $1\sigma$ statistical uncertainties, unless stated otherwise.

\section{Sample}\label{sec:sample}
We selected a sample of optically selected clusters based on the CAMIRA S20a v2 cluster catalog, an updated version of the HSC CAMIRA cluster catalog presented in \cite{Oguri18}.  Among the 21,250 optical clusters discovered in the Subaru HSC survey fields, 997 objects with richness $N>15$ and redshift $0.10<z<1.34$ lie in the eFEDS field. The richness range $N>15$ ($N>40$) corresponds to the cluster mass $M_{500}\gtrsim 5\times10^{13}~{\rm M_{\odot}}$ ($M_{500}\gtrsim 2\times10^{14}~{\rm M_{\odot}}$) at $z>0.1$ \citep{Okabe19}.  We cross-matched the CAMIRA catalog with the eFEDS X-ray cluster catalog \citep{Liu22} to find 211 optical clusters that have at least one spatially-extended X-ray source within the scale radius $R_{500}$ from the optical centers and the redshift difference of $|\Delta z| < 0.02$. Here, $R_{500}$ was estimated from the mass-richness relation \citep{Okabe19} and the cluster richness \citep{Oguri18}. Table~\ref{tab:xmatch} summarizes the result of the cross-matching of two catalogs.

\begin{table}[htb]
    \caption{Cross-matching of optical and X-ray cluster catalogs}\label{tab:xmatch}
    \centering
    \begin{tabular}{llll} \hline\hline
$\hat{N}_{\mathrm{mem}}$ & CAMIRA & eFEDS  & Fraction (\%) \\ \hline
15--20 & 524 & 27 & 5 \\
20--30 & 335 & 62 & 19 \\
30--40 & 97  & 50 & 52 \\
$>40$  & 41 & 32 & 78 \\ \hline
Total  & 997 & 171 & 17 \\ \hline
    \end{tabular}
\end{table}

In general, the X-ray detectability tends to be lower for lower richness and higher redshift; however, the observed fraction of 17\% is lower than expected from the known cluster mass-observable relations \citep{Okabe19} and eROSITA's sensitivity, when assuming the typical $\beta$-model brightness distribution of regular clusters \citep{Ota04}. This result can be attributed to multiple factors such as a large fraction of irregular clusters with low surface brightness \citep{Ota20}, the BCG miscentering effect \citep{Oguri18}, and misclassification as point sources \citep{Bulbul22}, thus requiring further investigation.

\begin{figure*}
    \centering
    \includegraphics[width=0.4\textwidth]{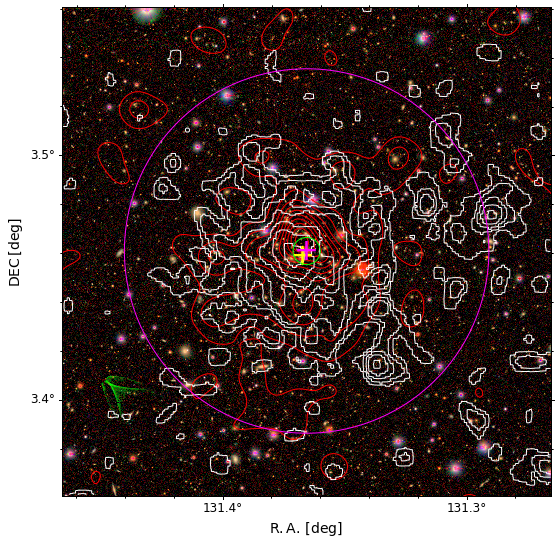}
    \includegraphics[width=0.4\textwidth]{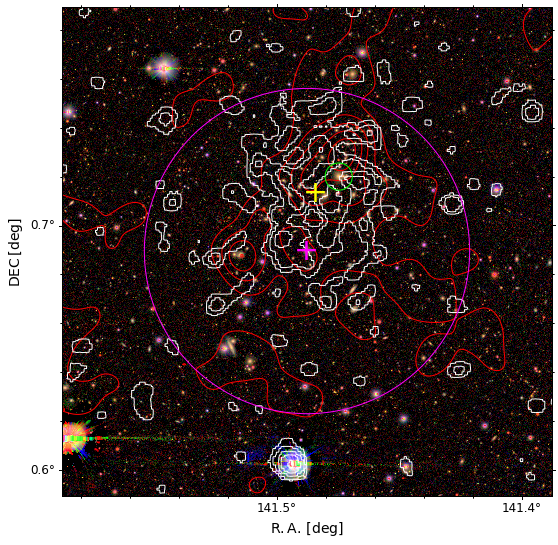}
    \caption{Examples of Subaru riz-composite images of the optically selected clusters with overlaid the eROSITA X-ray intensity maps, HSC~J084528+032739 (left) and HSC~J092557+004122 (right). The red contours are linearly spaced by half of the average height of galaxy density maps over all CAMIRA clusters at the same redshift. The white contours for X-ray emission are ten levels logarithmically spaced from $[10-1000]~{\rm cts\,s^{-1}deg^{-2}}$. In each panel, the X-ray centroid and the initial CAMIRA coordinates are marked by the yellow and pink cross, respectively. The green circle denotes the position of the confirmed BCG. The magenta circle indicates $R_{500}$.
 }\label{fig:hscxray}
\end{figure*}

To study the optical clusters' dynamical status and scaling relations based on the eROSITA analysis of individual objects, we focus on high-richness clusters with $N>40$ and $0.16<z<0.89$ in this paper. We note that there were 41 clusters with $N>40$ in the CAMIRA catalog (Table~\ref{tab:xmatch}); however, since several CAMIRA clusters have two galaxy-density peaks separated by $\gtrsim R_{500}$, we treat them as two individual clusters in the following analysis. There are seven clusters identified in this way: HSC~J085621+014649, HSC~J085629+014157, HSC~J092050+024514, HSC~J092246+034241, HSC~J093512+004738, HSC~J093501+005415, and HSC~J093523+023222. Their optical properties were obtained by the forced run of the CAMIRA optical cluster finder \citep{Klein22,Oguri18}. This yields a sample size of 43. Table~\ref{tab:sample} gives the sample list. Fig.~\ref{fig:hscxray} shows examples of HSC images with superposed X-ray contours. A statistical analysis of the whole cluster sample will be presented in a separate paper.

\begin{landscape}
\begin{table}
    \caption{Sample list.}\label{tab:sample}
    \centering
    \begin{tabular}{l l l l l l l l l } \hline\hline
    
    Cluster & $z$ & $\hat{N}_{\mathrm{mem}}$\tablefootmark{a} & $R_{500}$  & BCG position & X-ray centroid & X-ray peak & $D_{\mathrm{XC}}$\tablefootmark{b} & $D_{\mathrm{XP}}$\tablefootmark{c} \\
     & & & (Mpc/'') & RA, Dec (deg) & RA, Dec (deg) & RA, Dec (deg) & (kpc) & (kpc)\\ \hline

HSC J083655+025855 &     0.189 &  42.5 & 0.852 /          269 & 129.2144 ,   3.0004 &     129.2141 ,        2.9992 & 129.2163 ,    2.9988 &        13 &        27       (27,37) \\
HSC J083932-014128 &     0.255 &  45.7 & 0.866 /          217 & 129.8891 ,  -1.6791 &     129.8844 ,       -1.6831 & 129.8916 ,   -1.6790 &        87 &        35       (30,35) \\
HSC J084222+013826 &     0.418 &  75.8 & 1.050 /          188 & 130.5912 ,   1.6406 &     130.5975 ,        1.6463 & 130.5919 ,    1.6385 &       167 &        44      (44,657) \\
HSC J084245-000936 &     0.420 &  48.5 & 0.848 /          152 & 130.6896 ,  -0.1601 &     130.6991 ,       -0.1635 & 130.7021 ,   -0.1411 &       202 &       453     (453,453) \\
HSC J084435+031020 &     0.735 &  40.9 & 0.709 /           96 & 131.1453 ,   3.1723 &     131.1449 ,        3.1697 & 131.1443 ,    3.1703 &        68 &        57       (48,75) \\
HSC J084441+021656 &     0.649 &  71.3 & 0.949 /          135 & 131.1710 ,   2.2823 &     131.1582 ,        2.2847 & 131.1743 ,    2.2833 &       325 &        84       (65,94) \\
HSC J084502+012631 &     0.415 &  40.5 & 0.779 /          141 & 131.2596 ,   1.4420 &     131.2592 ,        1.4554 & 131.2533 ,    1.4626 &       265 &       426     (238,426) \\
HSC J084528+032739 &     0.320 & 103.6 & 1.255 /          268 & 131.3657 ,   3.4608 &     131.3675 ,        3.4589 & 131.3691 ,    3.4588 &        44 &        65       (65,69) \\
HSC J084548+020640 &     0.582 &  44.8 & 0.777 /          116 & 131.4498 ,   2.1054 &     131.4508 ,        2.1130 & 131.4619 ,    2.1203 &       181 &       454     (421,454) \\
HSC J084656+013836 &     0.597 &  41.1 & 0.743 /          110 & 131.7340 ,   1.6432 &     131.7481 ,        1.6482 & 131.7306 ,    1.6670 &       359 &       575     (575,644) \\
HSC J084824+041206 &     0.873 &  60.1 & 0.816 /          104 & 132.0988 ,   4.2017 &     132.0949 ,        4.2020 & 132.0995 ,    4.2023 &       107 &        25        (8,33) \\
HSC J084939-005121 &     0.616 &  53.4 & 0.836 /          122 & 132.4136 ,  -0.8560 &     132.4156 ,       -0.8572 & 132.4165 ,   -0.8583 &        57 &        91       (91,97) \\
HSC J085019+020011 &     0.419 &  44.8 & 0.817 /          146 & 132.5635 ,   2.0099 &     132.5762 ,        1.9969 & 132.5776 ,    2.0078 &       362 &       283     (275,284) \\
HSC J085056-000931 &     0.890 &  44.4 & 0.703 /           89 & 132.7330 ,  -0.1586 &     132.7328 ,       -0.1586 & 132.7301 ,   -0.1605 &         4 &        96      (86,103) \\
HSC J085232+002551 &     0.280 &  46.6 & 0.868 /          203 & 133.1256 ,   0.4205 &     133.1255 ,        0.4085 & 133.1270 ,    0.4066 &       183 &       213     (213,213) \\
HSC J085741+031135 &     0.188 &  73.6 & 1.108 /          351 & 134.4751 ,   3.1764 &     134.4669 ,        3.1810 & 134.4642 ,    3.1797 &       105 &       128     (125,128) \\
HSC J090143-014019 &     0.302 &  49.6 & 0.888 /          197 & 135.3779 ,  -1.6548 &     135.3775 ,       -1.6602 & 135.3715 ,   -1.6685 &        87 &       242      (88,441) \\
HSC J090330-013622 &     0.440 &  50.8 & 0.862 /          150 & 135.8756 ,  -1.6062 &     135.8718 ,       -1.6070 & 135.8570 ,   -1.6010 &        81 &       395     (389,395) \\
HSC J090451+033310 &     0.807 &  50.6 & 0.768 /          100 & 136.2176 ,   3.5537 &     136.2229 ,        3.5593 & 136.2183 ,    3.5606 &       207 &       189     (178,246) \\
HSC J090541+013226 &     0.636 &  40.4 & 0.727 /          105 & 136.4216 ,   1.5406 &     136.4212 ,        1.5366 & 136.4150 ,    1.5191 &        99 &       555      (26,555) \\
HSC J090754+005732 &     0.693 &  47.5 & 0.772 /          107 & 136.9765 ,   0.9590 &     136.9794 ,        0.9589 & 136.9765 ,    0.9561 &        74 &        72      (72,212) \\
HSC J090914-001220 &     0.305 &  50.7 & 0.897 /          198 & 137.3074 ,  -0.2056 &     137.3120 ,       -0.2067 & 137.3002 ,   -0.2079 &        77 &       122     (122,122) \\
HSC J090917-010134 &     0.823 &  49.9 & 0.758 /           98 & 137.3190 ,  -1.0261 &     137.3137 ,       -1.0191 & 137.3168 ,   -1.0188 &       239 &       205     (198,216) \\
HSC J091352-004535 &     0.264 &  49.1 & 0.893 /          218 & 138.4674 ,  -0.7597 &     138.4687 ,       -0.7580 & 138.4775 ,   -0.7622 &        31 &       151     (147,151) \\
HSC J091606-002338 &     0.297 & 113.9 & 1.322 /          297 & 139.0385 ,  -0.4045 &     139.0466 ,       -0.3970 & 139.0427 ,   -0.3974 &       176 &       131     (131,131) \\
HSC J091843+021231 &     0.268 &  57.6 & 0.964 /          233 & 139.7092 ,   2.2009 &     139.7056 ,        2.2142 & 139.7055 ,    2.2014 &       204 &        54       (54,60) \\
HSC J092024+013444 &     0.698 &  44.3 & 0.746 /          103 & 140.0981 ,   1.5788 &     140.0974 ,        1.5783 & 140.0990 ,    1.5784 &        21 &        27       (26,43) \\
HSC J092121+031713 &     0.345 &  95.4 & 1.198 /          243 & 140.3380 ,   3.2870 &     140.3392 ,        3.2891 & 140.3417 ,    3.2948 &        42 &       152     (152,152) \\
HSC J092211+034641 &     0.252 &  65.9 & 1.032 /          261 & 140.5319 ,   3.7663 &     140.5369 ,        3.7667 & 140.5467 ,    3.7763 &        71 &       252     (243,257) \\
HSC J092557+004122 &     0.260 &  57.8 & 0.967 /          239 & 141.4748 ,   0.7199 &     141.4845 ,        0.7133 & 141.4919 ,    0.7071 &       169 &       310     (310,310) \\
HSC J092847+005132 &     0.310 &  41.4 & 0.813 /          177 & 142.1952 ,   0.8587 &     142.2048 ,        0.8775 & 142.2499 ,    0.8913 &       345 &      1043    (540,1043) \\
HSC J092942+022843 &     0.537 &  43.6 & 0.778 /          121 & 142.4255 ,   2.5061 &     142.4232 ,        2.4956 & 142.4344 ,    2.5122 &       243 &       246      (75,246) \\
HSC J093025+021726 &     0.532 &  66.0 & 0.949 /          149 & 142.5929 ,   2.3068 &     142.6030 ,        2.2916 & 142.6045 ,    2.2907 &       414 &       450     (450,456) \\
HSC J093049-003651 &     0.310 &  43.4 & 0.832 /          181 & 142.7055 ,  -0.6159 &     142.7052 ,       -0.6197 & 142.7051 ,   -0.6175 &        62 &        27      (27,369) \\
HSC J093431-002308 &     0.342 &  45.5 & 0.842 /          172 & 143.6300 ,  -0.3855 &     143.6353 ,       -0.3870 & 143.6196 ,   -0.3844 &        96 &       183     (183,184) \\
HSC J094025+022840 &     0.164 &  66.4 & 1.062 /          377 & 145.1024 ,   2.4776 &     145.1188 ,        2.4822 & 145.1902 ,    2.4350 &       171 &       985     (520,985) \\
HSC J085621+014649 &     0.769 &  44.1 & 0.728 /           97 & 134.0858 ,   1.7804 &     134.0865 ,        1.7794 & 134.0886 ,    1.7810 &        34 &        77       (42,77) \\
HSC J085629+014157 &     0.692 &  42.6 & 0.733 /          102 & 134.1288 ,   1.7105 &     134.1294 ,        1.7077 & 134.1136 ,    1.7099 &        74 &       389     (388,389) \\
HSC J092050+024514 &     0.284 &  47.5 & 0.875 /          203 & 140.2071 ,   2.7539 &     140.2084 ,        2.7565 & 140.2067 ,    2.7506 &        45 &        50       (50,50) \\
HSC J092246+034241 &     0.257 &  55.3 & 0.948 /          236 & 140.6929 ,   3.7113 &     140.6893 ,        3.7079 & 140.6911 ,    3.7151 &        71 &        60      (58,133) \\
HSC J093512+004738 &     0.352 &  96.2 & 1.200 /          240 & 143.8018 ,   0.7939 &     143.8069 ,        0.7977 & 143.8083 ,    0.8010 &       112 &       172     (172,177) \\
HSC J093501+005415 &     0.374 &  53.9 & 0.905 /          174 & 143.7528 ,   0.9041 &     143.7553 ,        0.9036 & 143.7591 ,    0.9030 &        46 &       117     (109,117) \\
HSC J093523+023222 &     0.513 &  84.5 & 1.074 /          172 & 143.8439 ,   2.5569 &     143.8414 ,        2.5451 & 143.8383 ,    2.5436 &       267 &       321     (321,329) \\
\hline
    \end{tabular}
    \tablefoot{ 
\tablefoottext{a}{Richneess.}
\tablefoottext{b}{Centroid offset.}
\tablefoottext{c}{Peak offset. The error range estimated by changing the smoothing scale of the X-ray image is given in the parenthesis (see Sects.~\ref{subsec:centroid_determination} and \ref{subsec:morphology}). }
}
\end{table}
\end{landscape}

\section{Optical data analysis} \label{sec:optical}
\subsection{BCG identification} 
The CAMIRA algorithm \citep{Oguri14} first calculates the probability of each galaxy in the field being on the red sequence. Consequently, the galaxy clusters are defined as the overdensity of this type of galaxy population. At the same time, the selection method also returns the brightest cluster galaxy's (BCG) position for each cluster. This cluster's optical center is the position of the most luminous red-sequence galaxy close to the galaxy density peak. However, ultra-bright BCGs may not be correctly identified as BCGs because they are saturated in the HSC images, and the color estimates tend to be inaccurate. These reasons suggest that an additional confirmation of the BCG's position is necessary.

To determine the BCGs of all the clusters, we used the optical/NIR data from SDSS \citep{2000AJ....120.1579Y}, Pan-STARRS \citep{2002SPIE.4836..154K,2010SPIE.7733E..0EK}, 2MASS \citep{2006AJ....131.1163S}, and WISE \citep{2010AJ....140.1868W}. Each survey has a different sensitivity and filter response. In particular, SDSS has an i-band saturation level of 14 mag for point sources, which is brighter than the 18 mag of the HSC. In fact, for nearly half of the clusters, the brightest galaxies exceeded the HSC’s saturation level and were re-identified in the following way. We searched for the brightest galaxy that locates within an $R_{500}$ radius from the cluster's optical center, which is taken from the CAMIRA catalog. For the optical data, we used the $r$-band magnitudes to determine the brightest galaxy, whilst the $K_s$ band and $W_1$ values were used in the IR regime. The redshifts of the galaxies were either taken from the SDSS catalog or the NASA/IPAC Extragalactic Database (NED)\footnote{\url{https://ned.ipac.caltech.edu/}}. The redshift constraints of the BCGs search is $\Delta z=| z_{\mathrm{galaxy}}-z_{\mathrm{cluster}} | = 0.01\times (1+z_{\mathrm{cluster}})$ for the spectroscopic redshifts and $\Delta z = 0.02\times ({1+z_{\mathrm{cluster}}})$ for the photometric ones, respectively.  Here, we considered uncertainties of redshift measurements and redshift tolerance of $3\sigma/c \sim 0.01$ for the typical velocity dispersion of rich clusters, $\sigma\sim 1000~{\rm km\,s^{-1}}$ \citep{Fadda96}.

 Two additional corrections have been made before the photometric data are used for the BCGs search. First, we corrected for the galactic extinction, using the Schlegel map \citep{1998ApJ...500..525S}, assuming an extinction law \citep{1999PASP..111...63F} with $R_V$= 3.1. Second, appropriate $K$-correction methods are applied to obtain the magnitudes as in the rest frames of individual galaxies. We used the $K$-correction code version 2012 \citep{2010MNRAS.405.1409C,2012MNRAS.419.1727C} for the SDSS, Pan-STARRS, and 2MASS data because this method performs well with bright sources and does not require the compulsory input of multiple filter bands. However, this method does not provide the coefficients for WISE filters; therefore, we applied the $K$-correction code from \cite{2007AJ....133..734B} on the WISE photometric data instead.

Finally, we assessed the BCG search result through a careful visual inspection. In the case that multiple BCG candidates are suggested for one cluster, we assigned the BCG to the elliptical galaxy with a more extended envelope. If the search from different surveys returns different BCGs, we gave more weight to the BCG results from the optical observations. There were two cases (HSC~J091843+021231 and HSC~J092041+024660), where we needed to check the information of the obvious BCG candidates manually. These were the seemingly best choices but were not selected by the BCG finding program. Indeed, these galaxies do not have photometric imprints in any of the aforementioned surveys. We found the redshift of these visible BCG candidates on NED and their brightness in GAMA \citep{2016MNRAS.460..765W}. We thus confirmed that they are the BCGs of their host clusters because they also follow the spatial and redshift constraints.

\subsection{Weak-lensing mass measurement}\label{subsec:weaklens} 
Next, we describe weak-lensing (WL) analyses for the CAMIRA clusters in the eFEDS field. We used the latest shape catalog and the weak-lensing mass calibration from the three-year HSC data \citep[S19A;][]{Li22}. The galaxy shapes are measured by the re-Gaussianization method \citep{Hirata03} implemented in the HSC pipeline \citep[see details in][]{Mandelbaum18,HSCWL1styr}. The same shape catalog is used in the WL mass measurements for the eFEDS clusters \citep{Chiu22}. We adopted the full-color and full-depth criteria for precise shape measurements and photometric redshift estimations. 

The dimensional, reduced tangential shear, $\Delta \Sigma_{+}$, is computed by averaging the tangential component of a galaxy ellipticity $e_+=-(e_{1}\cos2\varphi+e_{2}\sin2\varphi)$ where $\varphi$ is the angle measured in sky coordinates from the RA direction to the line between the source galaxy and the lens. The formulation is specified by:
\begin{eqnarray}
\Delta \Sigma_{+} (R_k) = 
\frac{\sum_{i} e_{+,i} w_{i} \langle \Sigma_{{\rm cr}}(z_{l}, z_{s,i})^{-1}\rangle^{-1}}{2 \mathcal{R}(R_k) (1+K(R_k)) \sum_{i} w_{i}}, \label{eq:g+}
\end{eqnarray}
\citep[e.g.,][]{Miyaoka18,Medezinski18b,Okabe19,Miyatake19,Murata19,2020ApJ...890..148U,2021MNRAS.501.1701O,Chiu22}. Here, the subscripts, $i$ and $k$, denote the $i$-th galaxy located in the $k-$th radial bin, and $z_{l}$ and $z_{s}$ are the cluster and source redshift, respectively. The inverse of the mean critical surface mass density, $\langle \Sigma_{{\rm cr}}(z_{l},z_{s})^{-1}\rangle$, is calculated by weighting the critical surface mass density, $\Sigma_{{\rm cr}}=c^2D_{s}/4\pi G D_{l} D_{ls}$, by the probability function of the photometric redshift, $P(z)$: 
\begin{eqnarray}
 \langle \Sigma_{{\rm cr}}(z_{l},z_{s})^{-1}\rangle =
  \frac{\int^\infty_{z_{l}}\Sigma_{{\rm cr}}^{-1}(z_{l},z_{s})P(z_{s})dz_{s}}{\int^\infty_{0}P(z_{s})dz_{s}}, 
\end{eqnarray}
Here, $D_l$, $D_s$, and $D_{ls}$ are the angular diameter distances from the observer to the cluster, to the sources, and from the lens to the sources, respectively.  The photometric redshift is estimated by the machine learning method \citep[MLZ;][]{MLZ14} calibrated with spectroscopic data \citep{2020arXiv200301511N}. The dimensional weighting function is expressed as
\begin{eqnarray}
w=\frac{1}{e_{\rm rms}^2+\sigma_{e}^2}\langle \Sigma_{{\rm cr}}^{-1}\rangle^2,\label{eq:weight}
\end{eqnarray}
where $e_{\rm rms}$ and $\sigma_e$ are the root mean square of intrinsic ellipticity and the measurement error per component ($e_\alpha$; $\alpha=1$ or $2$), respectively. The shear responsivity, $\mathcal{R}$, and the calibration factor, $K$, are obtained by $\mathcal{R}=1-\sum_{ij} w_{i,j} e_{{\rm rms},i}^2/\sum_{ij} w_{i,j}$ and
$K=\sum_{ij} m_i w_{i,j}/\sum_{ij} w_{i,j}$, with the multiplicative
shear calibration factor $m$ \citep{Mandelbaum18,HSCWL1styr}, respectively.
We also conservatively subtracted an additional, negligible offset term for calibration.
The radius position, $R_k$, is defined by the weighted harmonic mean \citep{Okabe16b}.
We selected background galaxies behind each cluster using the photometric selection (p-cut) following \cite{Medezinski18}:
\begin{eqnarray}
  \int^\infty_{z_{l}+0.2}P(z)dz > p_{\rm cut},
\end{eqnarray}
where we allowed for a 2\% contamination level with $p_{\rm cut}=0.98$.

In order to measure individual cluster masses, we used the NFW profile \citep{NFW96,NFW97}. 
The three-dimensional mass density profile of the NFW profile is
expressed as 
\begin{equation}
\rho_{\rm NFW}(r)=\frac{\rho_s}{(r/r_s)(1+r/r_s)^2},
\label{eq:rho_nfw}
\end{equation}
where $r_s$ is the scale radius and $\rho_s$ is the central density
parameter. The NFW model is also specified by the spherical mass,
$M_{\Delta}=4\pi \Delta \rho_{\rm cr} r_{\Delta}^3/3$, and the halo
concentration, $c_{\Delta}=r_{\Delta}/r_s$. Here, $r_\Delta$ is the overdensity radius. 
We treat $M_\Delta$ and $c_\Delta$ as free parameters and adopt $\Delta=500$.
We computed the model of the dimensional, reduced tangential shear, $f_{\rm model}$, at the projected radius $R$ 
by integrating the mass density profile along the line of sight \citep{Okabe19,2020A&ARv..28....7U};
\begin{eqnarray}
f_{\rm model}(R)= \frac{\bar{\Sigma}(<R)-\Sigma(R)}{1-\mathcal{L}_z \Sigma(R)},
\end{eqnarray}
where $\Sigma(R)$ is the local surface mass density at $R$, 
$\bar{\Sigma}(<R)$ is the average surface mass density within $R$, 
and $\mathcal{L}_z=\sum_{i} \langle \Sigma_{{\rm cr},i}^{-1}\rangle w_{i}/\sum_{i} w_{i}$.
We chose the X-ray-defined centers and adopt an adaptive radial-bin choice \citep{Okabe16b} for cluster mass estimation. 

The log-likelihood of the weak-lensing analysis is expressed as 
\begin{eqnarray}
&& -2\ln {\mathcal L}_{\rm WL}=\ln(\det(C_{km})) +  \label{eq:likelihood}
  \\
&& \sum_{k,m}(\Delta \Sigma_{+,k} - f_{{\rm model}}(R_k))C_{km}^{-1} (\Delta
 \Sigma_{+,m} - f_{{\rm model}}(R_m)), \nonumber
\end{eqnarray}
where $k$ and $m$ denote the $k-$th and $m-$th radial bins.
We considered three components in the covariance matrix $C=C_g+C_s+C_{\rm LSS}$; the shape noise $C_g$, the errors of the source redshifts, $C_s$ and the uncorrelated large-scale structure (LSS), $C_{\rm LSS}$, of which the elements are correlated with each other \citep{Schneider98,Hoekstra03}. The details of calculations are described in Sect.~3 of \citet{Okabe16b}. We measured WL masses in 38 out of the 43 CAMIRA clusters that satisfy the full-color and full-depth conditions (Table~\ref{tab:spec}). The signal-to-noise ratios, $S/N=(\Delta \Sigma_{+,k}C_{km}^{-1}\Delta \Sigma_{+,m})^{1/2}$, of individual clusters were small; $S/N<2$ for 3, $2\le S/N<4$ for 20, and $4<S/N$ for 17 clusters, resulting in uncertainties in $M_{500}$. Here, for consistency with other X-ray observables, the center coordinates in the WL measurements were assumed to be equal to the X-ray centroids.

We also carried out the NFW model fitting with a free central position using two-dimensional shear pattern \citep{2010MNRAS.405.2215O,2011ApJ...741..116O}. 
The log-likelihood is expressed as
\begin{eqnarray}
  -2\ln\mathcal{L}_{\rm WL} = \sum_{\alpha,
   \beta=1}^2\sum_{k,m}\left[\Delta \Sigma_{\alpha,k} - f_{{\rm model},\alpha}\left(\bm{R}_k\right) \right]\bm{C}^{-1}_{\alpha\beta, km}\nonumber \\
    \times\left[ \Delta \Sigma_{\beta,m} - f_{{\rm model},\beta}\left(\bm{R}_m\right) \right]+\ln(\det(C_{\alpha\beta,km})).
\end{eqnarray}
Here, the subscripts $\alpha$ and $\beta$ denote each shear component.  
We used the box size of $1\arcmin.5 \times 1\arcmin.5$ for the shear pattern. We constrained the central positions for 23 clusters with good posterior distributions and compared them with X-ray centers, CAMIRA centers, and galaxy map peaks (Sect.~\ref{subsec:centroids}).

\section{X-ray data analysis} \label{sec:xray}
\subsection{Data reduction}
The eFEDs data of seven telescope modules (TMs) were reduced in a standard manner by using the eROSITA Standard Analysis Software System (eSASS) version {\tt eSASSusers\_201009} \citep{Brunner22}. We extracted cleaned event files by applying a flag to reject bad events and selecting all valid patterns, namely single, double, triple, and quadruple events. The point sources were removed by referring to the main eFEDS X-ray source catalog\footnote{\url{ https://erosita.mpe.mpg.de/edr/eROSITAObservations/Catalogues/}} \citep{Brunner22}. We checked that the faint sources with low detection likelihood in the supplemental eFEDS catalog do not affect our X-ray analysis. 

\subsection{Centroid and peak determination}\label{subsec:centroid_determination}
We extracted the X-ray image of each cluster from merged event files of seven TMs in the 0.5 -- 2~keV band and corrected them for exposure and vignetting. The pixel size is 1\arcsec.5. The X-ray centroid was determined from the mean of the photon distribution within a circle of radius $R_{500}$ using the algorithm described in \citet{Ota20}; Starting with the BCG coordinates, we iterated the centroid search until its position converged within 1\arcsec. The X-ray peak within $R_{500}$ was measured using the 0.5 -- 2~keV image smoothed with a $\sigma=3$ (pixels) Gaussian function. We note that we calculated the two-dimensional (2D) PSF image of the survey mode at each cluster coordinates using {\tt ermldet} in the eSASS package and confirmed that it is almost symmetric and does not affect the present measurement.

\subsection{Spectral analysis}
To measure the gas temperature and bolometric luminosity, we extracted the spectra from a circular region of a radius $R_{500}$ centered on the X-ray centroid. Here, it is reasonable to assume that the cluster center is represented by the X-ray centroid (Sect.~\ref{subsec:centroids}, Fig.~\ref{fig:2dwl}).  The TM1,2,3,4,6/TM5,7 spectra in the 0.3 -- 10~keV/1 -- 10~keV band were simultaneously fit using {\tt XSPEC 12.1.1}. For TM5 and TM7, the energies below 1~keV were excluded due to the light-leak contamination \citep{Predehl21}. The spectral model consists of cluster emission and background components. For the cluster component, we assumed the APEC thin-thermal plasma model \citep{Smith01, Foster12}, with the Galactic absorption model {\tt tbabs} \citep{Wilms00}. The redshift and metal abundance were fixed at the optical value (Table~\ref{tab:sample}) and 0.3~solar, respectively.  The hydrogen column density $N_{\rm H}$ in the {\tt tbabs} model was fixed at a value taken from \citet{Willingale13}. For the background components, the Galactic emission and cosmic X-ray background were determined by fitting an annulus with inner and outer radii of 2.5 and 4~Mpc from the cluster. The instrumental background was estimated based on the filter wheel closed (FWC) data\footnote{The FWC spectral model version 1.0  (\url{https://erosita.mpe.mpg.de/edr/eROSITAObservations/EDRFWC/})}.  Table~\ref{tab:spec} lists the resultant gas temperature, $kT,$ and bolometric luminosity, $L_{\rm X}$.  We note that for three clusters (HSC~J084548+020640, HSC~J084656+01383, and HSC~J090754+005732), we fixed $kT$ at a value expected from the $N-T$ relation \citep{Oguri18} and the richness due to the large statistical uncertainty. 

\begin{table*}[htb]
    \caption{Results of X-ray spectral analysis and lensing mass measurements}\label{tab:spec}
    \centering
    \begin{tabular}{lllll} \hline\hline
Cluster &  $N_{\rm H}$ & $kT$\tablefootmark{a} & $L_{\rm X}$\tablefootmark{b}  & $M_{500}$\tablefootmark{c}  \\
        & ($10^{20}~{\rm cm^{-2}}$) & (keV) & ($10^{44}~{\rm erg\,s^{-1}}$) & ($10^{14}~{\rm M_{\odot}}$) \\ \hline
HSC~J083655+025855 & 3.54 & $ 2.39 ^{+0.36 }_{ -0.28 } $ & $ 1.31 ^{+0.09 }_{ -0.08 } $ & $ 4.57 ^{+1.70 }_{ -1.43 } $ \\ 
HSC~J083932-014128 & 3.09 & $ 3.71 ^{+0.71 }_{ -0.55 } $ & $ 3.73 ^{+0.30 }_{ -0.26 } $ & -- \\ 
HSC~J084222+013826 & 5.52 & $ 1.95 ^{+0.48 }_{ -0.35 } $ & $ 2.55 ^{+0.25 }_{ -0.25 } $ & $ 9.28 ^{+3.05 }_{ -2.57 } $ \\ 
HSC~J084245-000936 & 3.33 & $ 3.34 ^{+3.04 }_{ -1.17 } $ & $ 1.33 ^{+0.41 }_{ -0.24 } $ & $ 5.66 ^{+2.51 }_{ -2.00 } $ \\ 
HSC~J084435+031020 & 3.71 & $ 2.37 ^{+1.21 }_{ -0.53 } $ & $ 3.43 ^{+0.53 }_{ -0.44 } $ & $ 7.27 ^{+9.67 }_{ -7.26 } $ \\ 
HSC~J084441+021656 & 5.17 & $ 2.42 ^{+0.66 }_{ -0.45 } $ & $ 5.98 ^{+0.60 }_{ -0.55 } $ & $ 15.92 ^{+18.92 }_{ -9.18 } $ \\ 
HSC~J084502+012631 & 4.58 & $ 4.37 ^{+3.40 }_{ -1.54 } $ & $ 2.18 ^{+0.57 }_{ -0.37 } $ & $ 2.67 ^{+1.63 }_{ -1.25 } $ \\ 
HSC~J084528+032739 & 3.63 & $ 7.72 ^{+3.52 }_{ -1.43 } $ & $ 8.69 ^{+1.18 }_{ -0.75 } $ & $ 5.32 ^{+1.64 }_{ -1.37 } $ \\ 
HSC~J084548+020640 & 4.82 & $ 3.69 $(fix) & $ 1.45 ^{+0.30 }_{ -0.28 } $ & $ 2.69 ^{+3.16 }_{ -2.26 } $ \\ 
HSC~J084656+013836 & 4.20 & $ 3.54 $(fix) & $ 0.44 ^{+0.22 }_{ -0.21 } $ & $ 5.22 ^{+4.78 }_{ -2.83 } $ \\ 
HSC~J084824+041206 & 4.12 & $ 4.09 ^{+3.58 }_{ -1.44 } $ & $ 6.15 ^{+1.67 }_{ -0.94 } $ & -- \\ 
HSC~J084939-005121 & 2.60 & $ 6.29 ^{+10.99 }_{ -2.96 } $ & $ 5.02 ^{+2.17 }_{ -1.20 } $ & $ 2.81 ^{+3.33 }_{ -2.87 } $ \\ 
HSC~J085019+020011 & 3.24 & $ 2.76 ^{+4.47 }_{ -1.12 } $ & $ 0.89 ^{+0.41 }_{ -0.20 } $ & $ 1.68 ^{+1.52 }_{ -1.27 } $ \\ 
HSC~J085056-000931 & 3.21 & $ 1.44 ^{+0.57 }_{ -0.24 } $ & $ 3.25 ^{+0.69 }_{ -0.61 } $ & $ 7.37 ^{+15.94 }_{ -10.67 } $ \\ 
HSC~J085232+002551 & 4.13 & $ 2.67 ^{+1.27 }_{ -0.64 } $ & $ 1.11 ^{+0.20 }_{ -0.14 } $ & $ 2.69 ^{+1.17 }_{ -0.91 } $ \\ 
HSC~J085741+031135 & 4.15 & $ 5.94 ^{+1.30 }_{ -0.88 } $ & $ 5.25 ^{+0.43 }_{ -0.35 } $ & $ 3.21 ^{+1.57 }_{ -1.29 } $ \\ 
HSC~J090143-014019 & 2.28 & $ 5.39 ^{+2.37 }_{ -1.31 } $ & $ 5.00 ^{+0.81 }_{ -0.58 } $ & -- \\ 
HSC~J090330-013622 & 2.26 & $ 5.00 ^{+7.49 }_{ -1.90 } $ & $ 3.37 ^{+1.43 }_{ -0.62 } $ & -- \\ 
HSC~J090451+033310 & 3.60 & $ 5.06 ^{+4.35 }_{ -1.69 } $ & $ 6.51 ^{+1.80 }_{ -1.00 } $ & -- \\ 
HSC~J090541+013226 & 4.01 & $ 3.88 ^{+2.90 }_{ -1.56 } $ & $ 2.90 ^{+0.72 }_{ -0.53 } $ & -- \\ 
HSC~J090754+005732 & 3.39 & $ 3.80 $ (fix) & $ 3.14 ^{+0.43 }_{ -0.41 } $ & $ 0.16 ^{+3.12 }_{ -2.04 } $ \\ 
HSC~J090914-001220 & 3.00 & $ 3.22 ^{+0.97 }_{ -0.77 } $ & $ 1.82 ^{+0.23 }_{ -0.20 } $ & $ 1.99 ^{+1.32 }_{ -0.93 } $ \\ 
HSC~J090917-010134 & 2.84 & $ 2.98 ^{+1.61 }_{ -0.72 } $ & $ 5.93 ^{+0.99 }_{ -0.70 } $ & -- \\ 
HSC~J091352-004535 & 3.20 & $ 2.21 ^{+0.84 }_{ -0.44 } $ & $ 0.85 ^{+0.12 }_{ -0.10 } $ & $ 3.50 ^{+1.17 }_{ -1.00 } $ \\ 
HSC~J091606-002338 & 3.20 & $ 4.39 ^{+1.10 }_{ -0.79 } $ & $ 5.40 ^{+0.56 }_{ -0.45 } $ & $ 8.65 ^{+3.26 }_{ -2.52 } $ \\ 
HSC~J091843+021231 & 2.74 & $ 2.57 ^{+0.49 }_{ -0.34 } $ & $ 2.12 ^{+0.17 }_{ -0.15 } $ & $ 2.80 ^{+1.47 }_{ -1.09 } $ \\ 
HSC~J092024+013444 & 2.90 & $ 2.95 ^{+0.70 }_{ -0.43 } $ & $ 10.09 ^{+0.82 }_{ -0.66 } $ & $ 10.84 ^{+8.61 }_{ -6.84 } $ \\ 
HSC~J092121+031713 & 3.87 & $ 6.19 ^{+1.52 }_{ -1.09 } $ & $ 10.38 ^{+0.96 }_{ -0.82 } $ & $ 5.12 ^{+1.69 }_{ -1.37 } $ \\ 
HSC~J092211+034641 & 3.98 & $ 5.33 ^{+1.34 }_{ -1.06 } $ & $ 4.53 ^{+0.45 }_{ -0.41 } $ & $ 3.82 ^{+1.52 }_{ -1.27 } $ \\ 
HSC~J092557+004122 & 3.43 & $ 3.25 ^{+1.27 }_{ -0.90 } $ & $ 1.09 ^{+0.17 }_{ -0.15 } $ & $ 2.03 ^{+1.26 }_{ -0.98 } $ \\ 
HSC~J092847+005132 & 4.27 & $ 1.74 ^{+0.57 }_{ -0.27 } $ & $ 0.46 ^{+0.08 }_{ -0.07 } $ & $ 1.08 ^{+1.04 }_{ -0.86 } $ \\ 
HSC~J092942+022843 & 4.85 & $ 3.53 ^{+4.22 }_{ -1.35 } $ & $ 1.69 ^{+0.61 }_{ -0.33 } $ & $ 1.86 ^{+1.95 }_{ -1.38 } $ \\ 
HSC~J093025+021726 & 5.06 & $ 6.98 ^{+12.14 }_{ -1.97 } $ & $ 8.27 ^{+3.30 }_{ -1.09 } $ & $ 5.11 ^{+3.40 }_{ -2.45 } $ \\ 
HSC~J093049-003651 & 3.09 & $ 2.43 ^{+0.73 }_{ -0.40 } $ & $ 1.16 ^{+0.15 }_{ -0.11 } $ & $ 3.21 ^{+1.42 }_{ -1.19 } $ \\ 
HSC~J093431-002308 & 3.15 & $ 2.78 ^{+1.23 }_{ -0.66 } $ & $ 1.59 ^{+0.26 }_{ -0.19 } $ & $ 9.13 ^{+4.05 }_{ -2.83 } $ \\ 
HSC~J094025+022840 & 3.72 & $ 2.18 ^{+1.29 }_{ -0.52 } $ & $ 1.31 ^{+0.30 }_{ -0.21 } $ & $ 7.10 ^{+2.21 }_{ -1.75 } $ \\ 
HSC~J085621+014649 & 3.87 & $ 4.49 ^{+1.60 }_{ -0.91 } $ & $ 15.93 ^{+2.06 }_{ -1.42 } $ & -- \\ 
HSC~J085629+014157 & 3.97 & $ 3.21 ^{+1.06 }_{ -0.69 } $ & $ 6.62 ^{+0.78 }_{ -0.65 } $ & $ 1.46 ^{+4.00 }_{ -3.20 } $ \\ 
HSC~J092050+024514 & 3.50 & $ 3.05 ^{+0.56 }_{ -0.42 } $ & $ 2.84 ^{+0.22 }_{ -0.19 } $ & $ 2.84 ^{+2.65 }_{ -1.38 } $ \\ 
HSC~J092246+034241 & 3.93 & $ 1.61 ^{+0.38 }_{ -0.25 } $ & $ 0.42 ^{+0.06 }_{ -0.06 } $ & $ 1.13 ^{+0.93 }_{ -0.78 } $ \\ 
HSC~J093512+004738 & 4.34 & $ 5.90 ^{+1.77 }_{ -1.12 } $ & $ 10.40 ^{+1.15 }_{ -0.89 } $ & $ 12.39 ^{+3.74 }_{ -2.89 } $ \\ 
HSC~J093501+005415 & 4.46 & $ 3.12 ^{+1.07 }_{ -0.61 } $ & $ 2.73 ^{+0.36 }_{ -0.26 } $ & $ 3.16 ^{+1.52 }_{ -1.38 } $ \\ 
HSC~J093523+023222 & 3.43 & $ 3.95 ^{+0.78 }_{ -0.56 } $ & $ 14.71 ^{+1.13 }_{ -0.92 } $ & $ 5.41 ^{+3.50 }_{ -3.50 } $ \\ \hline
    \end{tabular}
    \tablefoot{ 
\tablefoottext{a}{Gas temperature;}
\tablefoottext{b}{Bolometric luminosity within the scale radius $R_{500}$}
; \tablefoottext{c}{ Weak-lensing mass within $R_{500}$. }
}
\end{table*}

\section{Results} \label{sec:results}
\subsection{Centroid offsets} \label{subsec:centroids} 
\begin{figure*}[htb]
     \centering
     \includegraphics[width=0.4\textwidth]{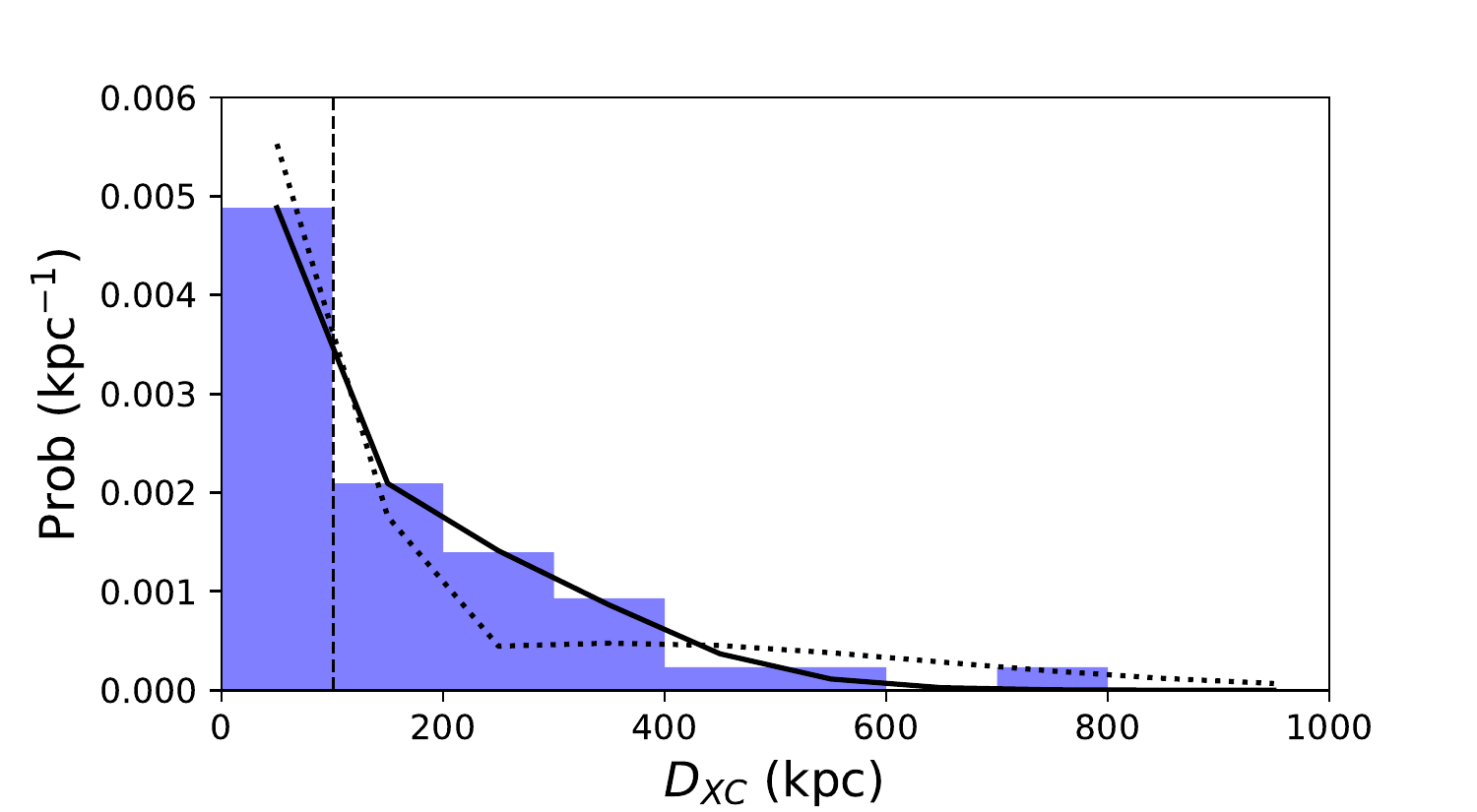}
     \includegraphics[width=0.4\textwidth]{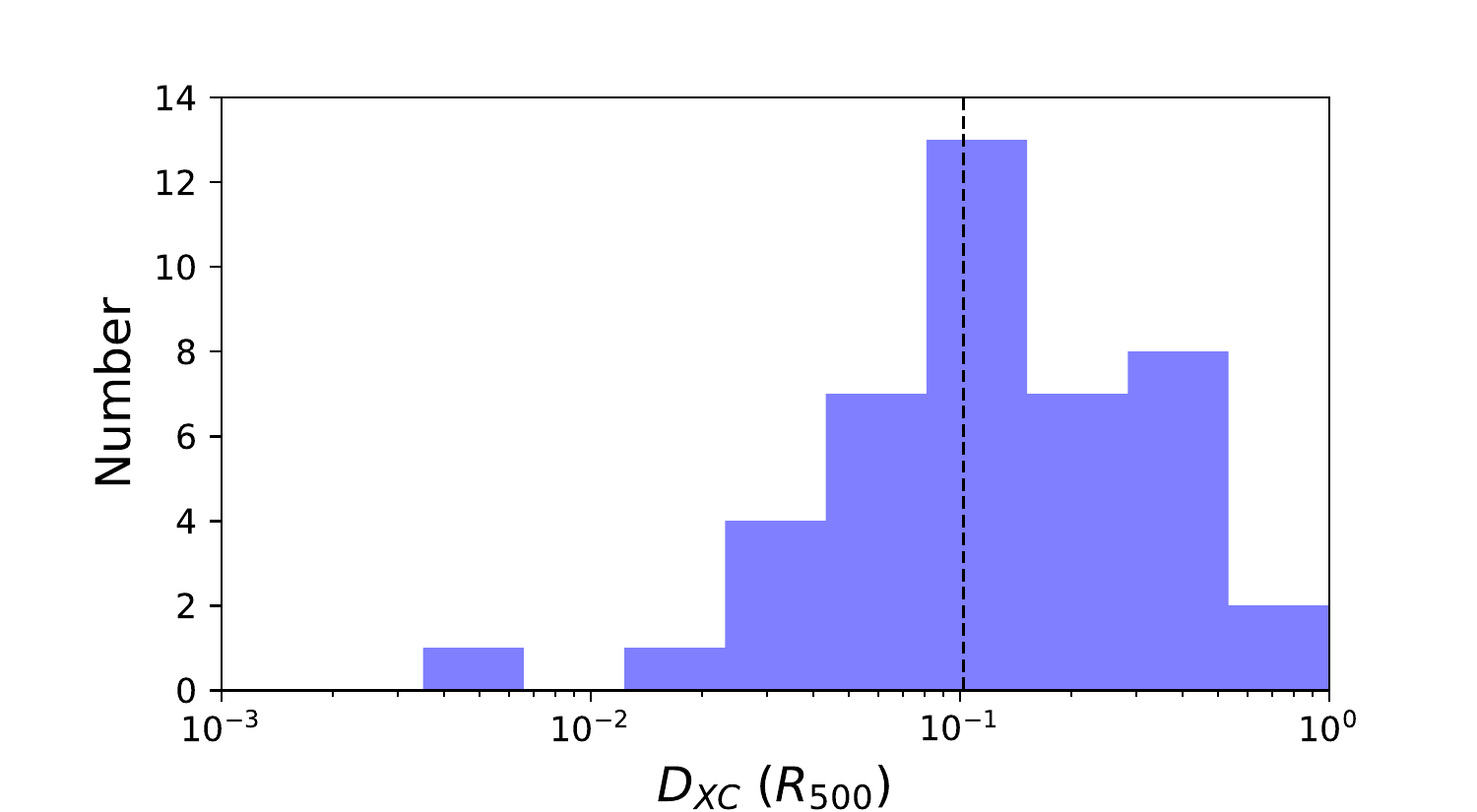}
     \includegraphics[width=0.4\textwidth]{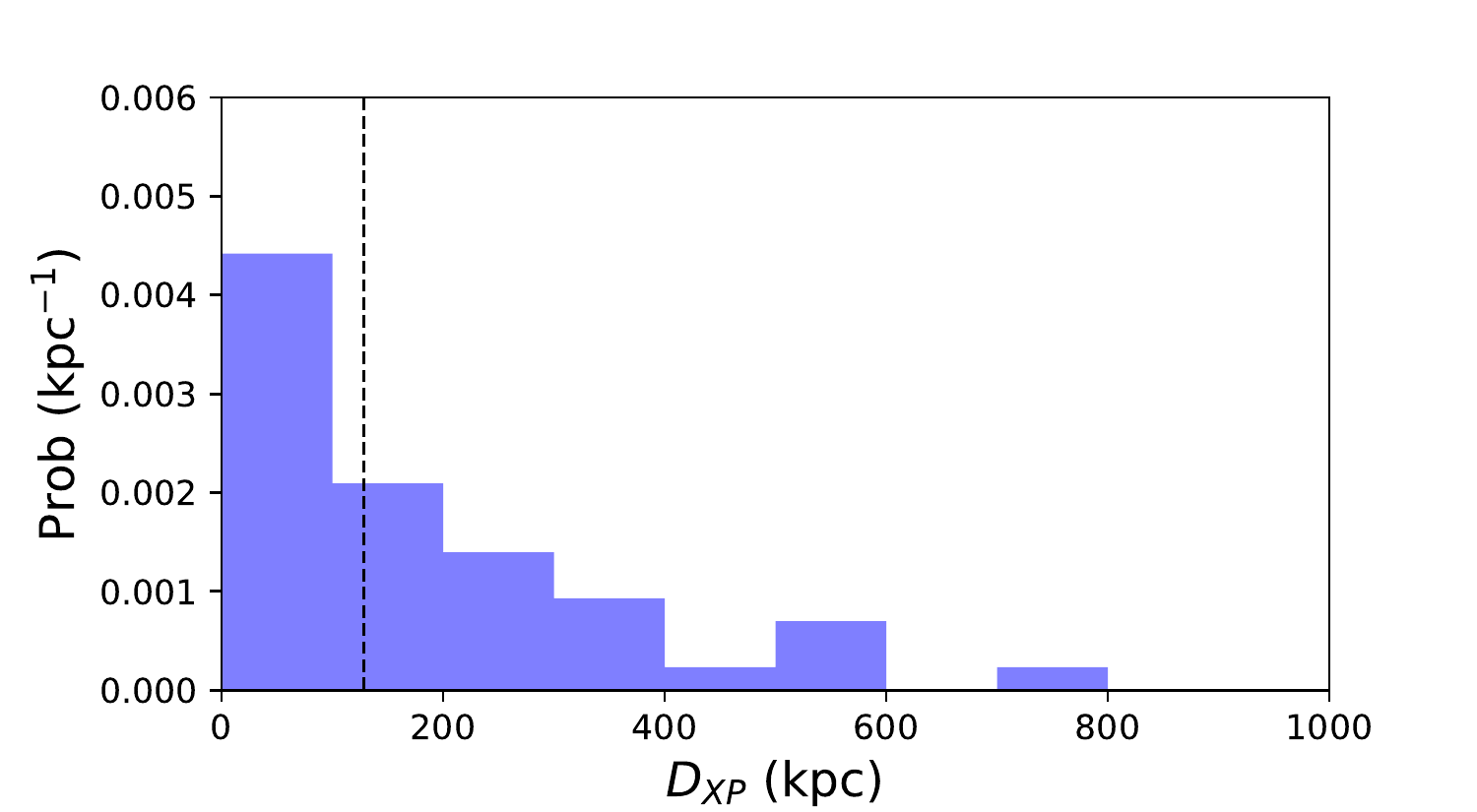}
     \includegraphics[width=0.4\textwidth]{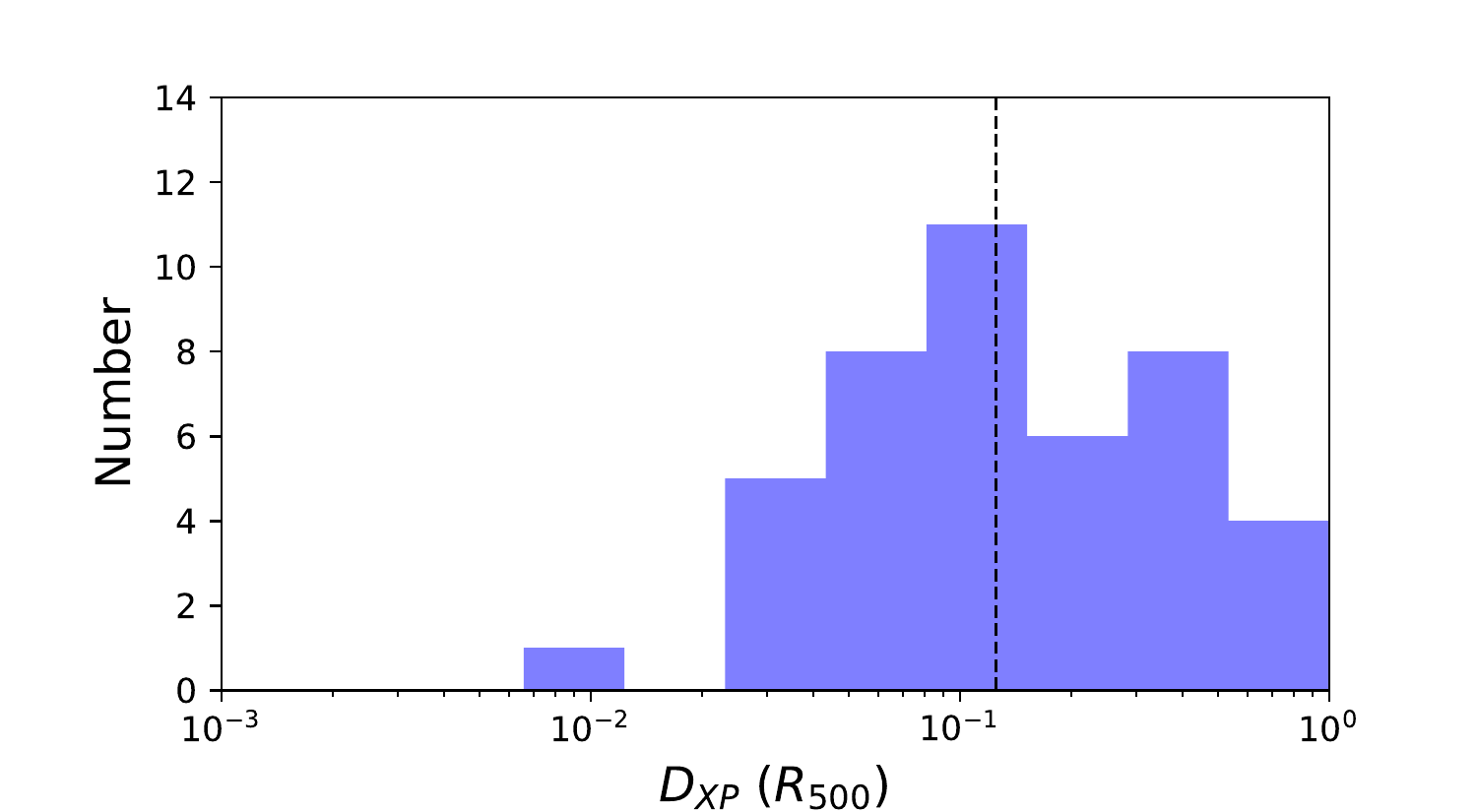}    
     \caption{Distributions of centroid offset (upper panels) and peak offset (lower panels) in units of kpc (left panels) and $R_{500}$ (right panels). In each panel, the dashed line indicates the median. In the upper-left panel, the solid and dotted curves show the best-fit double-Gaussian models for the present sample and the positional offset between the CAMIRA and XXL clusters \citep{Oguri18}, respectively.}\label{fig:histogram}
 \end{figure*}

We calculated the spatial separation from the cluster's X-ray centroid to the BCG's position and report it as the BCG-and-X-ray centroid offset, $D_{\mathrm{XC}}$. In addition, we estimate the distance between the cluster's X-ray peak to its BCG $D_{\mathrm{XP}}$. Figure~\ref{fig:histogram} demonstrates the distributions of $D_{\rm XC}$ and $D_{\rm XP}$ for 43 clusters in our sample. The median values of these offsets in the unit of kpc and $R_{500}$ are shown in Table~\ref{tab:median}. 

\begin{table}[htb]
    \caption{Median values of centroid offset and peak offset}\label{tab:median}
    \centering
    \begin{tabular}{lllll} \hline\hline
 $D_{\mathrm{XC}}$ & $D_{\mathrm{XC}}$ & $D_{\mathrm{XP}}$ & $D_{\mathrm{XP}}$ & Fraction\tablefootmark{a}\\ 
 (kpc) & ($R_{\mathrm{500}}$) & (kpc) & ($R_{\mathrm{500}}$)&  \\ \hline
100 & 0.10 & 152 & 0.14 & 2(<16)\% \\ \hline
    \end{tabular}
\tablefoot{
\tablefoottext{a}{Fraction of relaxed clusters based on the BCG-X-ray peak offset (Sect.~\ref{subsec:morphology}).}
}
\end{table}

\begin{figure*}
    \centering
    \includegraphics[width=0.33\textwidth]{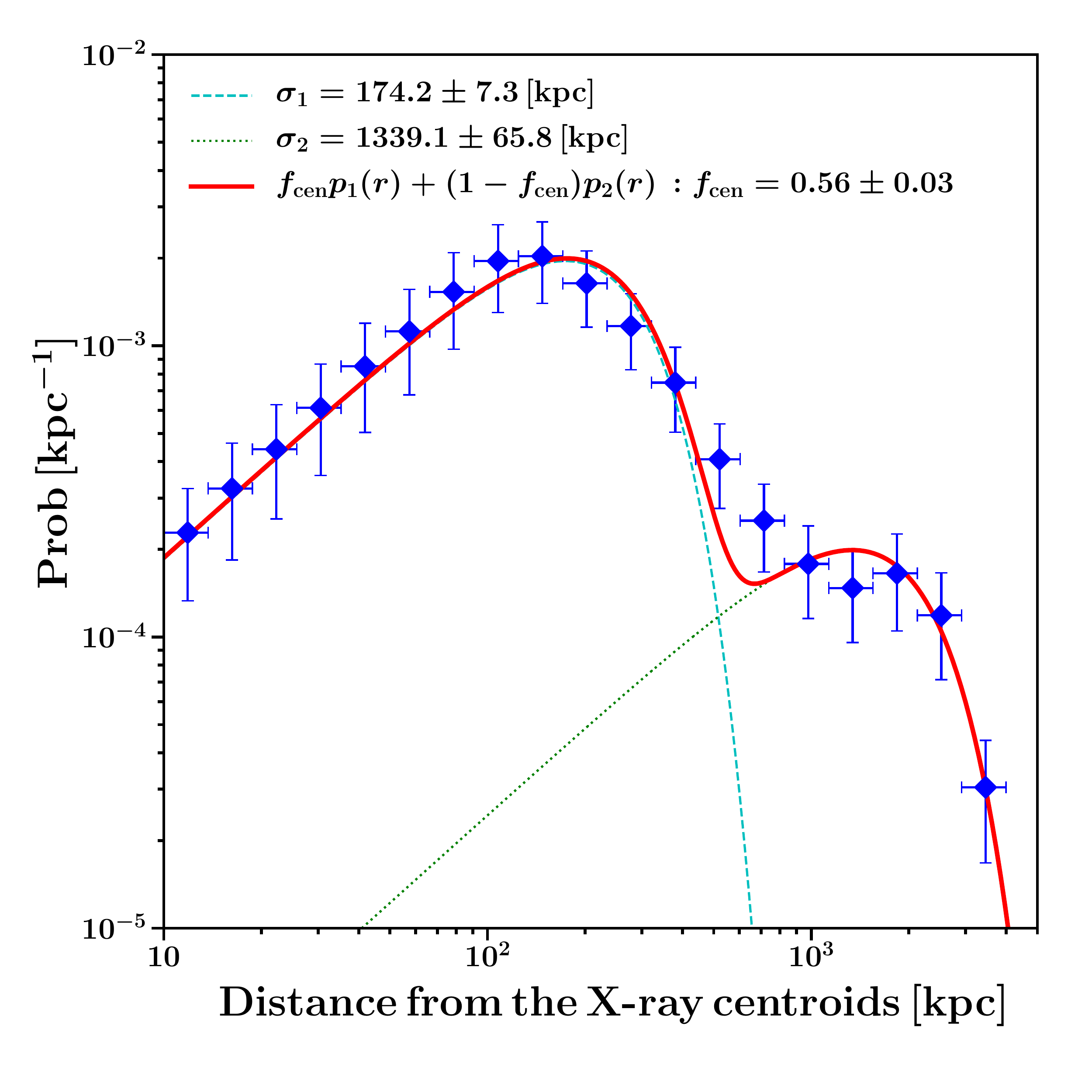}    
    \includegraphics[width=0.33\textwidth]{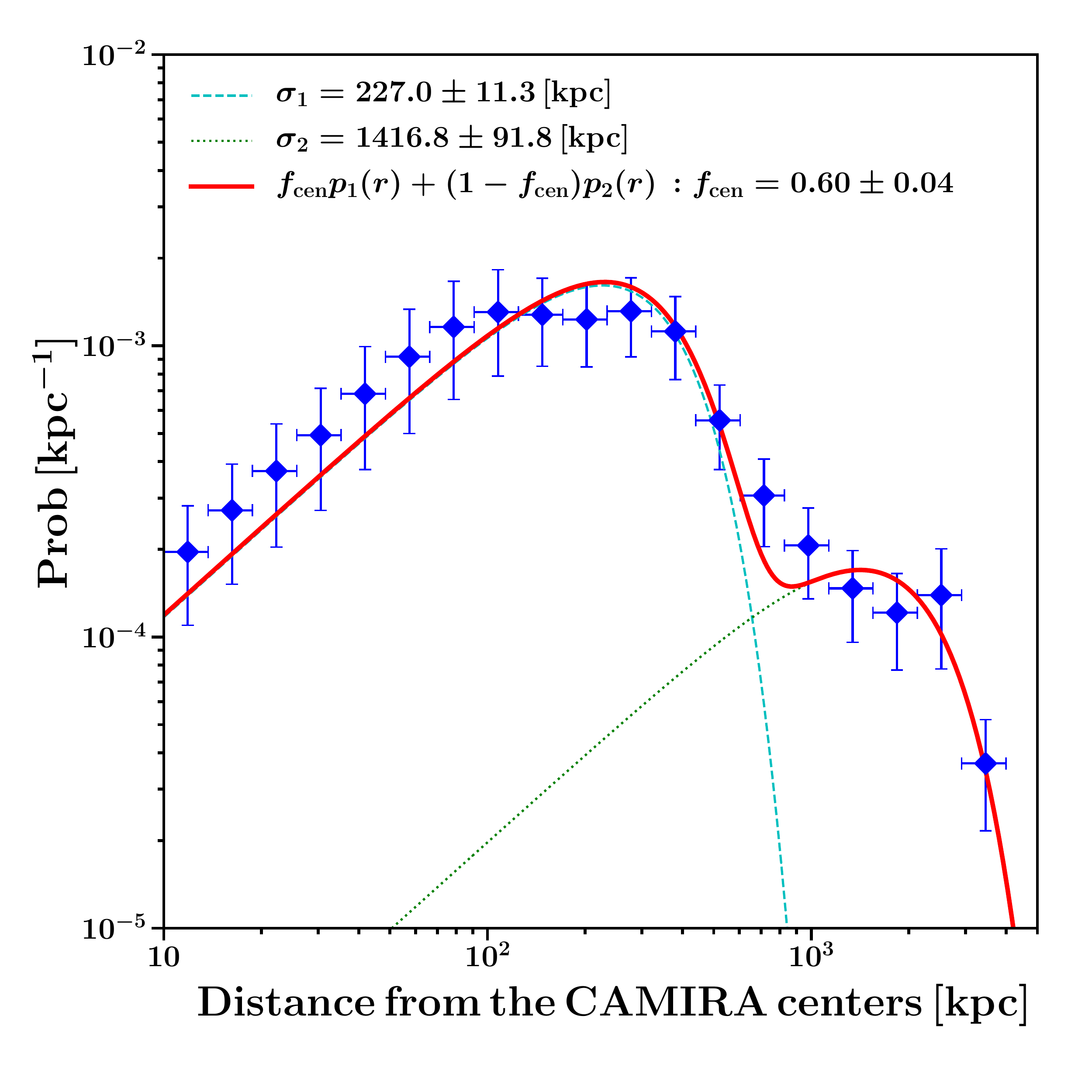}    \includegraphics[width=0.33\textwidth]{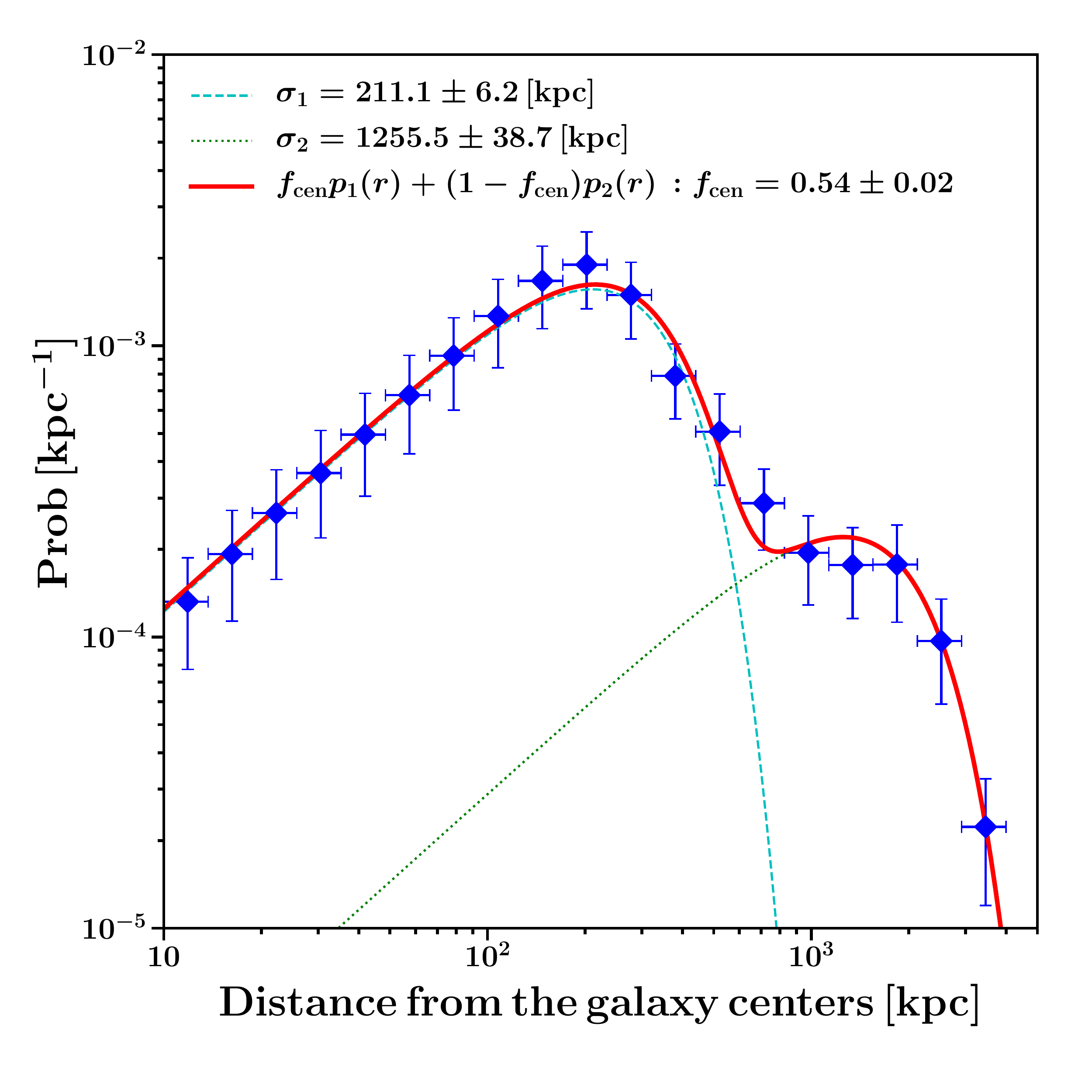}
     \caption{Histograms of the projected distance between 2D weak-lensing mass center and X-ray centroid (left), CAMIRA center (middle), galaxy map peak (right) for 23 CAMIRA clusters. The error bars indicate the statistical uncertainties. The best-fit model (red) consisting of the first and second Gaussian components is indicated by the cyan dashed, and green dotted curves, respectively.}\label{fig:2dwl}
\end{figure*}

Next, we compare the projected distance between 2D weak-lensing mass and three kinds of cluster centers, namely X-ray centroid, galaxy map peak, and the CAMIRA coordinates for 23 clusters. We estimated the statistical errors of the probability distributions by the bootstrap resampling method.  As shown in Fig.~\ref{fig:2dwl}, all the histograms are well described by double Gaussian distributions:
\begin{equation}
p(r)=f_{\rm cen} p_1(r)+(1-f_{\rm cen}) p_2(r), \label{eq:doublegauss}
\end{equation}
where $p_i(r)=(r/\sigma_i^2)\exp\left[-r^2/2\sigma_i^2\right]$ and $f_{\rm cen}$ is the fraction of the central component \citep{2010MNRAS.405.2215O,Oguri18}.
The first standard deviation, $\sigma_1$, for the X-ray centroid is smaller than those of the CAMIRA and galaxy ones, suggesting that the X-ray centroids are closer to the WL mass centers. Although the values of $\sigma_1$ for the CAMIRA centers are the largest and the histogram is slightly skewed from the double Gaussian distributions, the amplitude of the CAMIRA centers at $r\simlt 60$~kpc is comparable to that of the X-ray centers. Although the value of $\sigma_1$ for the galaxy peak is the second smallest among the first components of the three centers, the amplitude at $r\simlt 100$ kpc is $\sim60-70$ \% of the others. The second standard deviation, $\sigma_2$, of the galaxy peak is the smallest among the three centers.

\subsection{Morphological classification}\label{subsec:morphology}
In this subsection, we use three different methods to classify the morphology of galaxy clusters: i) BCG-X-ray offset, ii) concentration parameter, and iii) galaxy peak-finding method.

First, the BCG-X-ray offset: We divided the clusters into two groups following the criteria used in \cite{Sanderson09}, namely, ``relaxed'' clusters with a small peak offset ($D_{\rm XP}<0.02R_{500}$) and ``disturbed'' clusters with a large offset ($D_{\rm XP}>0.02R_{500}$). 
There is only one cluster that posseses a small offset of $D_{\mathrm{XP}} \leq 0.02 R_{500}$, which corresponds to a very small fraction of relaxed clusters $2.3\pm 2.2\,(^{+0.0}_{-2.3})$\% or < 5\%. Here, the statistical and systematic errors are calculated via the bootstrap resampling method and by varying the smoothing scales between 2 and 4 pixels, respectively.

The accuracy of the X-ray peak position depends on the statistical quality of the X-ray observations, which broadly varies among clusters. In fact, based on the XMM-Newton observations of the CAMIRA
clusters, \cite{Ota20} reported that the clusters with low photon statistics tend to show larger centroid and peak offsets. We thus calculated the standard error of the peak offset $\delta D_{\rm XP}$ based on a comparison of X-ray images of each cluster with different smoothing scales ($\sigma=2,3,4$~pixels). As a result, $\delta D_{\rm XP}$ ranges from 0\% to 55 \% (the mean is 5\%), which is smaller than that of the XMM-Newton study \citep{Ota20}. 

At high redshifts, however, $0.02R_{500}$ is comparable to the accuracy of the attitude determination \citep{Predehl21}. With regard to how much the results would change based on the threshold, when we set the threshold value to $0.05R_{500}$, the percentage of relaxed clusters was $9$\%. This example suggests that the systematic uncertainty of the relaxed fraction is not negligible. Accordingly, we assigned the typical positional accuracy of 4\arcsec.7 \citep{Brunner22} to the systematic error of $D_{\rm XP}$. This increases the number (or fraction) of relaxed clusters that fall in $D_{\mathrm{XP}} \leq 0.02 R_{500}$ to 7 (or 16\%). Consequently, we estimate the fraction of relaxed clusters as $2(<16)$\% from the BCG-X-ray offset (Table~\ref{tab:median}).

We should note that this modest fraction of relaxed clusters is potentially impacted by other effects. First, the selection of the BCG itself faces undeniable obstacles. The lack of redshift information and the use of photometric redshifts with large errors could lead to a poor choice of BCG. In some cases, the true BCG may not be detected due to being overbright for the optical instruments.

Second, there is the concentration parameter: 
if we refer to the concentration parameter of the X-ray surface brightness, $C_{{\rm SB}, R_{500}}$, and apply $C_{{\rm SB}, R_{500}}>0.37$ to identify relaxed clusters \citep{Lovisari17}, the fraction of relaxed clusters is estimated as 39\% for 33 optical clusters with X-ray counterparts in the eFEDS catalog within $R_{500}$ from the X-ray centroid and $\Delta z<0.02$. Here, $C_{{\rm SB}, R_{500}}$ is defined as $C_{{\rm SB}, R_{500}} =S_B(<0.1R_{500})/S_B(<R_{500})$ \citep{Maughan12} and we quoted the value measured by \cite{Ghirardini22}, based on the X-ray morphological study of the eFEDS clusters. Because all ten unmatched clusters are disturbed clusters, according to the BCG-X-ray offset measurements, we consider the above estimate to be an upper limit on the relaxed fraction. 

Third, the galaxy peak-finding method: 
Since our optically selected sample covers a wide range of morphologies, we expect a large percentage of merging clusters. These complex systems could be classified using the peak-finding method \citep{Okabe19}. Thus, we also checked the fraction of merging clusters by finding peaks of member-galaxy distribution \citep{Ramos-Ceja22}. The threshold corresponds to the peak height of the richness $N=15$ at each redshift. In the eFEDS field, the fraction of clusters with single over multiple peaks is 27/73\% for the high-richness CAMIRA clusters, while the fraction of single over multiple peaks is 83/17\% for the eFEDS clusters. 

To be conservative, we quote the results from three kinds of measurements and estimate the relaxed fraction of our sample to be $2(<39)$\%.

\subsection{Scaling relations}\label{subsec:scaling}
\begin{figure*}
     \centering
     \includegraphics[width=0.48\textwidth]{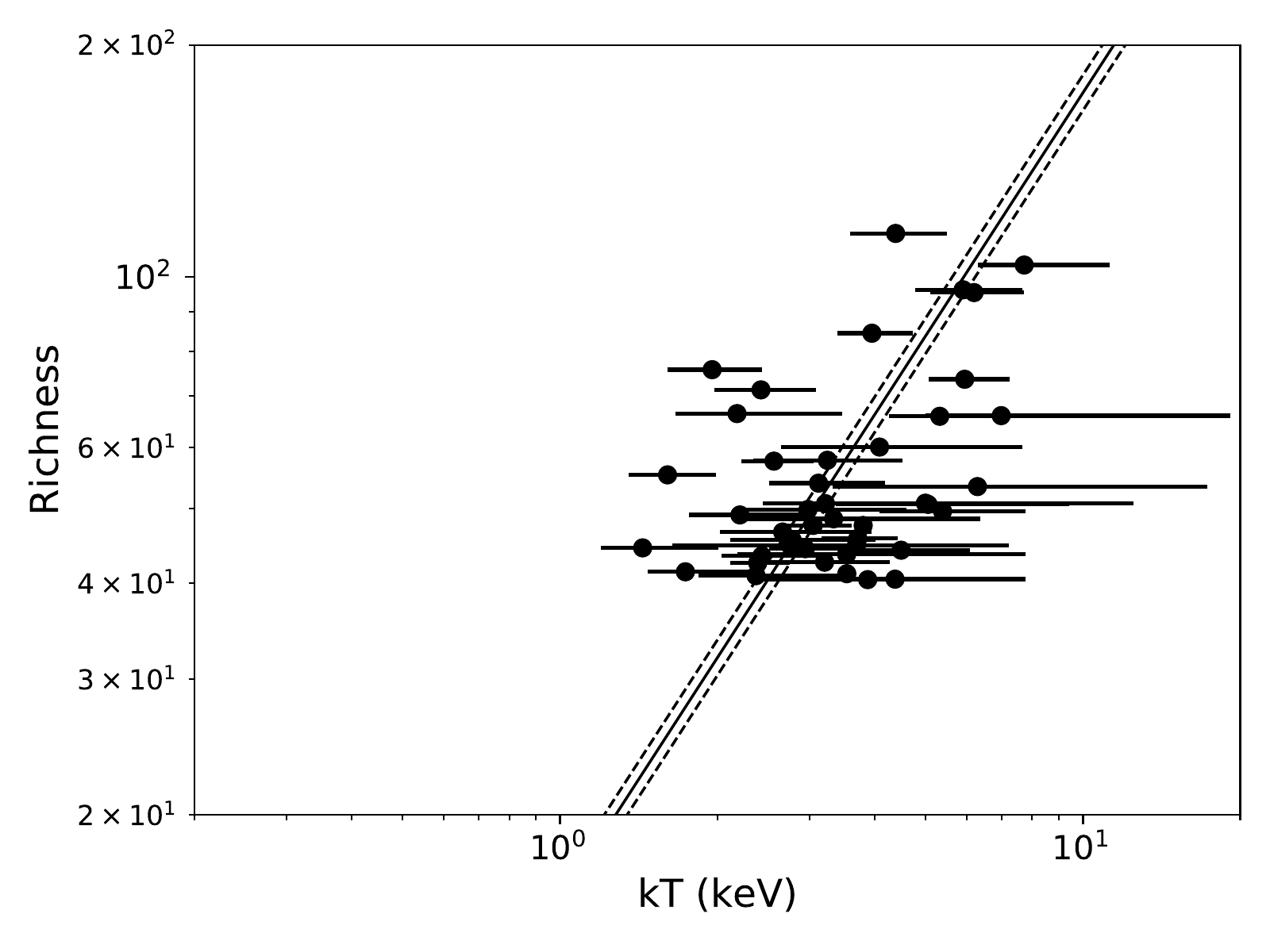}
     \includegraphics[width=0.48\textwidth]{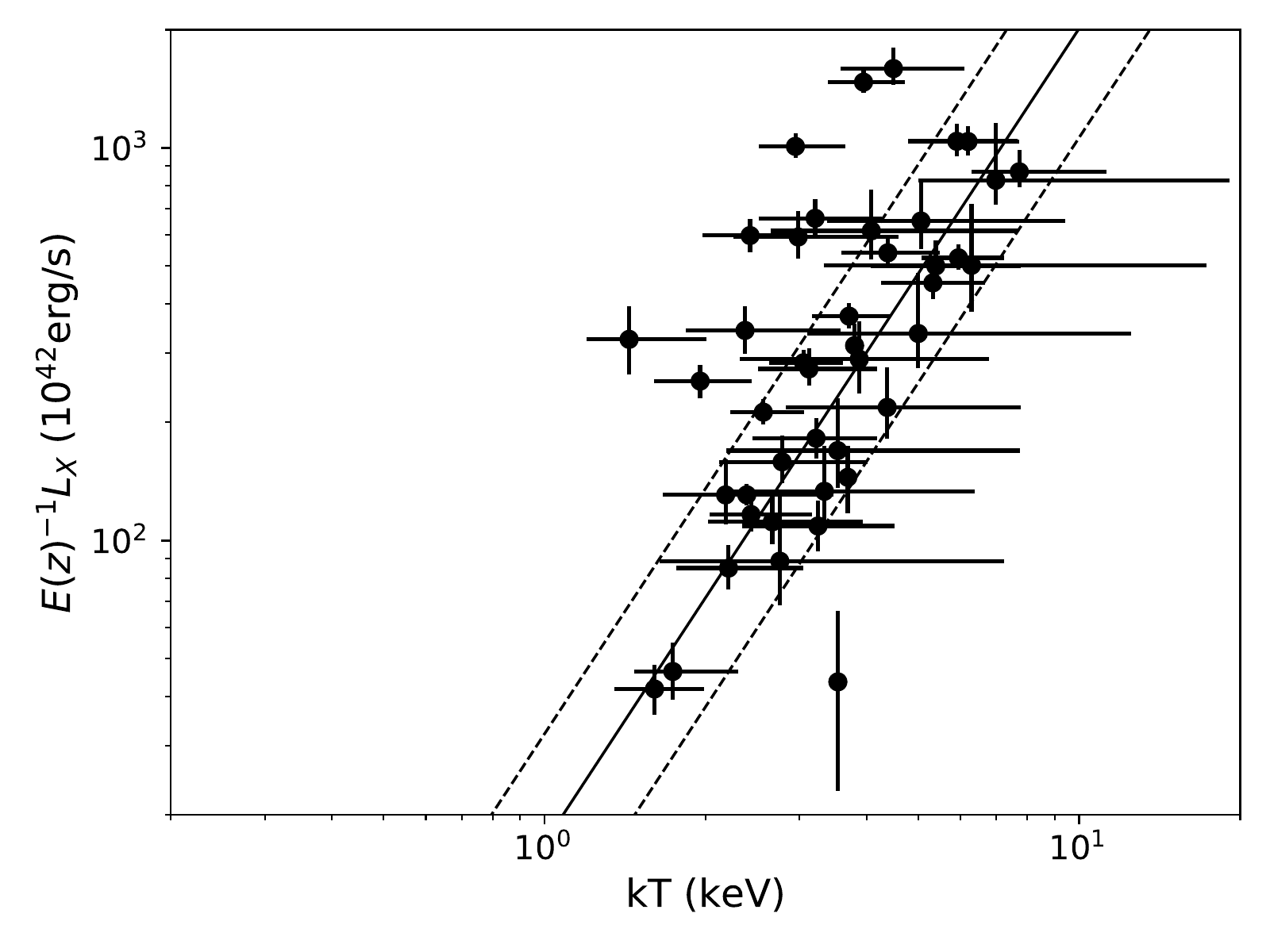}
     \includegraphics[width=0.48\textwidth]{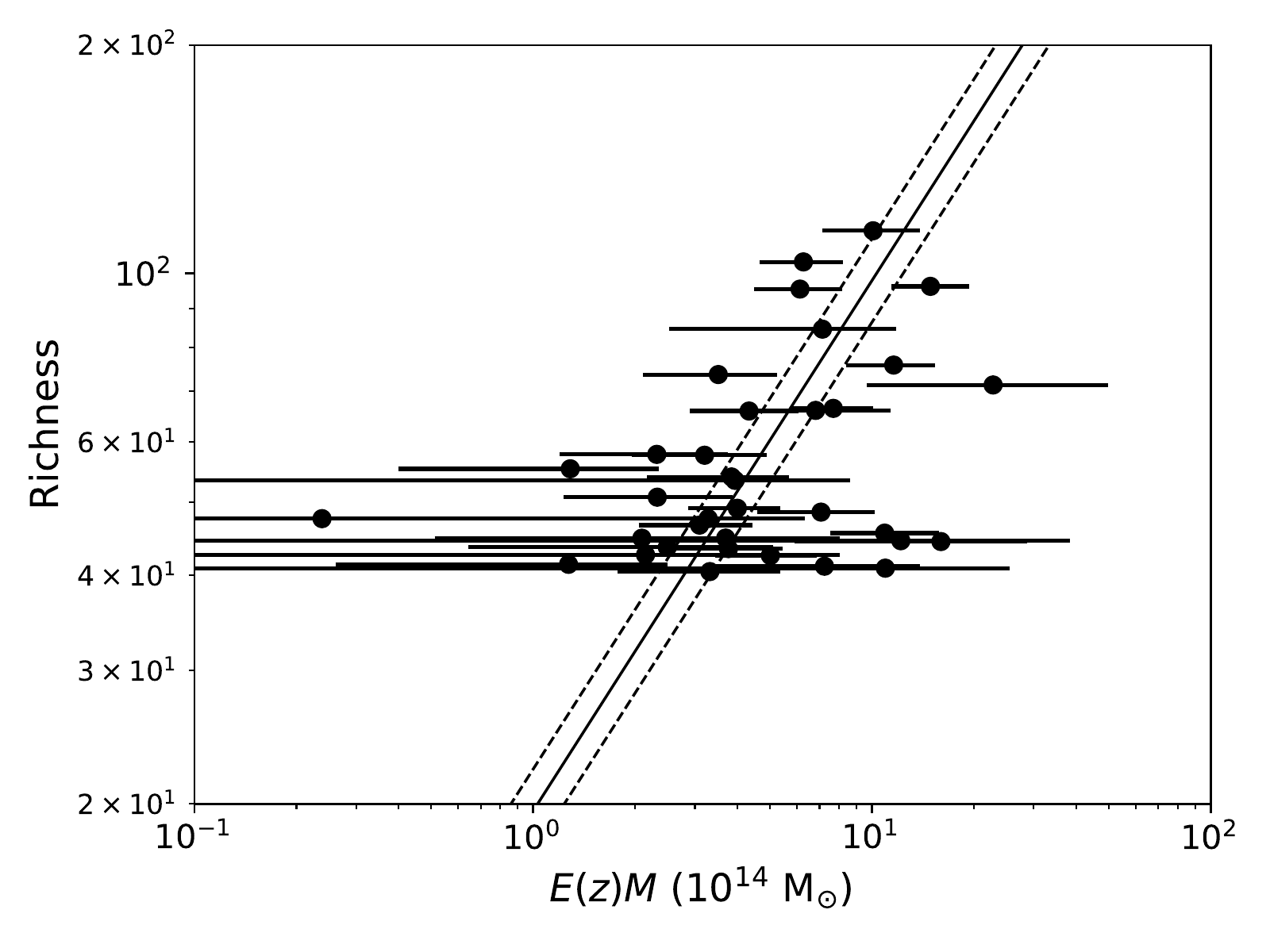}
     \includegraphics[width=0.48\textwidth]{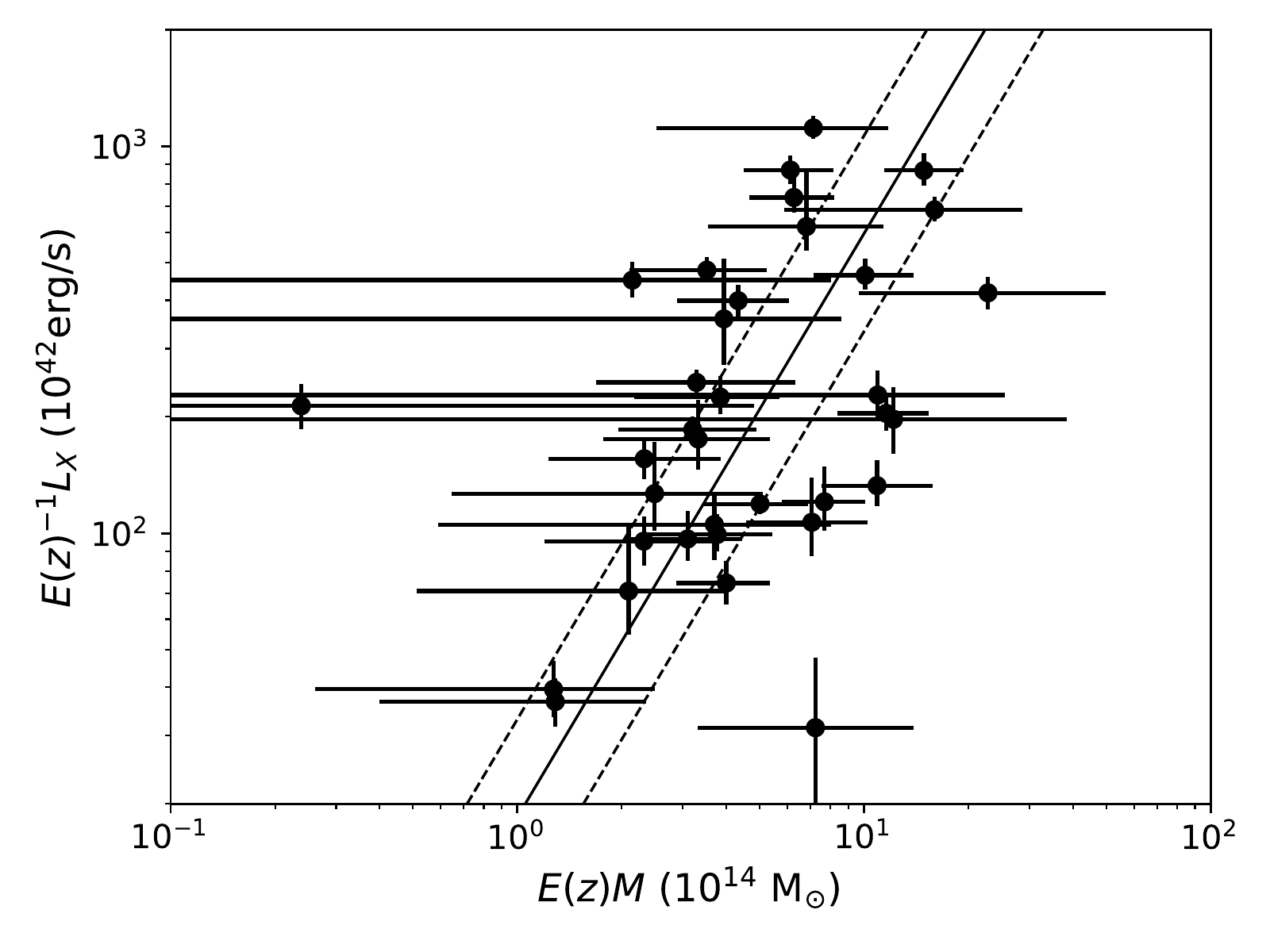}
    \caption{Scaling relations of the high-richness, optically selected clusters (circles). In the upper panels, the simultaneous fit of $T-N$ and $T-L$ relations are shown. In the lower panels, $M-N$ and $M-L$ relations are shown. In each panel, the best-fit power-law model and the $1\sigma$ uncertainty around the mean relation are indicated with the solid and dashed lines, respectively. }\label{fig:scaling}
 \end{figure*}

To derive the temperature- and mass-observable relations of the high-richness clusters, we fit the data to the power-law model (Eq.~\ref{eq:model}) via the Bayesian regression method \citep{Kelly07}: 
\begin{equation}
\log{y} = a + b\log{x}.  \label{eq:model}
\end{equation}
The quantities $a$, $b$, and the intrinsic scatter are treated as free parameters. According to the self-similar model and Hubble's law, $E(z)=[\Omega_M(1+z)^3+\Omega_{\Lambda}]^{1/2}$ describes the redshift evolution of the scaling relation. The luminosity of the cluster in the hydrostatic equilibrium follows $E(z)^{-1}L\propto T^2$ and $E(z)^{-1}L\propto [E(z)M]^{4/3}$. In Fig.~\ref{fig:scaling}, we corrected the redshift evolution by applying the self-similar model and plotted $E(z)^{-1}L$ against $T$ or $E(z)M$ since no clear consensus has been reached on the evolution of the scaling relations \citep{Giodini13}.  Table~\ref{tab:scaling} lists the best-fit parameters for the $N-T$ and $L-T$ relations of 43 clusters and the $N-M$ and $L-M$ relations for 38 clusters with the weak-lensing measurements (Sect.~\ref{subsec:weaklens}). The correlation coefficient is 0.62 -- 0.70 for the four relations. 

As described in \cite{Akino22}, a multi-variate analysis is needed to properly correct the selection bias and the dilution effect and incorporate the weak-lensing mass calibration. Thus, to correct for the selection bias due to the richness cut of $N>40$, we simultaneously fit two kinds of relations, $N-T$ and $L-T$ or $N-M$ and $L-M$, by the hierarchical Bayesian regression method \citep[HiBRECS;][]{Akino22}  (Fig.~\ref{fig:scaling}). Because the slopes of $N-T$ and $N-M$ were not well constrained due to the large data scatter, we fixed them at 1.05 and 0.70,
respectively, that were deduced from the best-fit mass-richness relation \citep{Okabe19}. We also assumed the intrinsic scatter of the weak-lensing mass to be $\ln{\sigma_M}=-1.54$ \citep{2020ApJ...890..148U}. Table~\ref{tab:scaling} summarizes the best-fit scaling relations obtained from three types of fitting codes.

From Table~\ref{tab:scaling}, we find that the fitting results with the Kelly and 1D HiBRECS codes agree well. The Kelly code gave a marginally shallower $N-T$ slope of $0.61\pm0.21, $ as compared to the expectation from $M\propto T^{3/2}$ and $M\propto N^{1.4}$\citep{Okabe19}.  In the present analysis, the impact of selection bias correction was small; the 1D and 2D HiBRECS analyses gave consistent results within the error bars. Therefore, in Sect.~\ref{sec:discussion}, we quote the above results based on the 2D HiBRECS code, which can properly handle the bias correction and the mass calibration.

\begin{table*}[htb]
    \caption{Best-fit scaling relations of the optically selected clusters obtained from three types of fitting codes}\label{tab:scaling}
    \centering
    \begin{tabular}{llllllllllll} \hline\hline
         & \multicolumn{3}{c}{Kelly} & \multicolumn{3}{c}{1D HiBRECS} & \multicolumn{3}{c}{2D HiBRECS} \\
Relation & $a$ & $b$ & $\sigma$ & $a$ & $b$  & $\sigma$ & $a$ & $b$  & $\sigma$ \\ \hline
$N-T$          & $1.44\pm0.10$ & $0.61\pm0.21$ & 0.095 & $1.19\pm0.02$ & 1.05 (fix)   & 0.24 & $1.07\pm0.04$ & 1.05 (fix) & 0.30 (fix) \\ 
$E(z)^{-1}L-T$ & $1.38\pm0.24$ & $1.87\pm0.45$ & 0.28  & $1.48\pm0.21$ & $1.70\pm0.40$ & 0.63 & $1.23\pm0.28$ & $2.08\pm0.46$ & 0.63 \\ \hline 
$N-E(z)M$      & $1.43\pm0.19$ & $0.45\pm0.24$ & 0.10  & $1.22\pm0.03$ & 0.70 (fix)  & 0.25 &  $1.29\pm0.06$ & 0.70 (fix) & 0.16  \\
$E(z)^{-1}L-E(z)M$ & $1.44\pm0.32$ & $1.22\pm0.43$ & 0.32 & $1.50\pm0.29$ & $1.27\pm0.25$ & 0.65 & $1.24\pm0.25$ & $1.52\pm0.34$ & 0.65 \\
\hline
$N-T$ & -- & -- & -- & -- &  -- &  -- & $1.07\pm0.03$ & 1.05 (fix) & 0.30 (fix) \\
$L-T$ & $1.45\pm0.27$ & $1.94\pm0.50$ & 0.31 & $1.57\pm0.24$ & $1.72\pm0.44$ & 0.68 & $1.41\pm0.30$ & $1.97\pm0.50$  & 0.70 \\ \hline
$N-M$ & $1.36\pm0.22$ & $0.61\pm0.33$ & 0.094& $1.31\pm0.03$ & 0.70 (fix) & 0.11 & $1.24\pm0.08$ & 0.70 (fix) & 0.27 \\
$L-M$ & $1.56\pm0.34$ & $1.33\pm0.49$ & 0.33 & $1.62\pm0.30$ & $1.24\pm0.45$ & 0.72 & $1.39\pm0.32$ & $1.58\pm0.44$ & 0.64 \\
\hline
\end{tabular}
\tablefoot{$T$, $L$, $M$ are in the units of keV, $10^{42}~{\rm erg\,s^{-1}}$, $10^{14}M_{\odot}$, respectively. The best-fit values of $a$, $b$ in Eq.~\ref{eq:model}, and the intrinsic scatter of $y$ are listed. The results with and without $E(z)$-correction are shown in 1--4 and 5--8 rows, respectively. The 2D HiBRECS code takes into account the correction for the selection bias and mass calibration, but the other two codes do not (Sects.~\ref{subsec:scaling} and \ref{subsec:discussion_scaling}).}
\end{table*}

\section{Discussion} \label{sec:discussion}
\subsection{BCG-X-ray offset and cluster dynamical status} \label{subsec:dynamical}
As mentioned in Sect.~\ref{sec:results}, only $2  (<16)$\% clusters out of our cluster sample possess a small X-ray peak offset $D_{\mathrm{XC}} \leq 0.02 R_{500}$. Compared with the previous XMM-Newton measurements of 17 optical clusters with $N>20$, $29 \pm 11 (\pm 13)$\% \citep{Ota20}, the two results agree within the errors. 

\cite{2016MNRAS.457.4515R} investigated the dynamical states of 132 galaxy clusters of a Sunyaev-Zel’dovich (SZ) selected a sample at a median redshift of $z = 0.16$ to find that $52\pm 4$\% of the sample has $D_{\mathrm{XP}}\leq$ $0.02R_{500}$. The SZ selection method tends to pick out dynamically-active systems, which leads to a relatively lower percentage of small peak shift. On the other hand, the relaxed ratio is higher in the X-ray selected sample, at $\approx$ 74\%, indicating that clusters collected in an X-ray flux-limited survey are subject to the cool-core bias. This accounts for a large fraction of X-ray bright, hence, dynamically relaxed clusters with a small peak shift. In addition, \cite{Migkas2021} found that approximately 44\% of the X-ray flux-limited eeHIFLUGCS sample have a small peak shift of $D_{XP}<0.02R_{500}$. Therefore, our sample's relaxed fraction is considerably small compared to the above SZ and X-ray samples. We note that \cite{Ghirardini22} reported that the eFEDS-selected cluster sample is not biased toward cool-core clusters, but that it does contain a similar fraction of cool-cores as SZ surveys.

As shown in the upper-left panel of Fig.~\ref{fig:histogram}, we assessed the distribution of the centroid offset by the double Gaussian model (Eq.~\ref{eq:doublegauss}) since there is a tail toward large $D_{\rm XC}$. The fitting yields $\sigma_1=28\pm2$~kpc, $\sigma_2=95\pm6$~kpc, and $f_{\rm cen}=0.46\pm0.04$. In comparison with the positional offset between the CAMIRA and XXL clusters \citep{Oguri18}, the present sample has a marginally smaller fraction of well-centered clusters and a lower tail component.

\cite{Pasini22} showed that BCGs with radio-loud active galactic nucleus (AGN) are more likely to lie close to the cluster center than radio-quiet BCGs and that the relations between the AGN and the intracluster medium (ICM) hold regardless of the dynamical state of the cluster \citep[see also][]{2020MNRAS.497.2163P}.

\subsection{Scaling relations} \label{subsec:discussion_scaling}
In what follows, we discuss the $L-T$ and $L-M$ relations based on a comparison with previous observations and theoretical models. The $L-T$ relation shows a large intrinsic scatter of $\sigma_{L|T}=0.63,$ as noted by the previous X-ray studies \citep[e.g.,][]{Ota06,Pratt09}. 
 The observed $L-T$ slope of  $2.08\pm0.46$ agrees with $2.2\pm0.6\,(\pm 0.2)$ derived from the XMM-Newton observations of the CAMIRA clusters \citep{Ota20}. In contrast, a steeper slope of $\sim 3$ has been reported by many X-ray observations in the past \citep{Giodini13}. From the analysis of 265 eFEDS clusters, \cite{Bahar22} obtained  the best-fit $L_{\rm bol}-T$ slope of $3.01^{+0.13}_{-0.12}$, suggesting a strong deviation from the self-similar model, $L \propto T^2$. Because of the small fraction of relaxed clusters, we consider the present measurement is less affected by the cool-core emission. 

The scatter of $L-M$ relation is comparably large, $\sigma_{L|M}=0.65$, which confirms previous reports \citep[e.g.,][]{Pratt09}. The $L-M$ slope of $1.52\pm0.34$ is consistent with $1.51\pm0.09$ derived for 232 clusters at $z=0.05-1.46$ based on a compilation of 14 published X-ray data sets \citep{Reichert11}. The best-fit $L-M$ relation of the present optical sample also agrees with the result on 25 shear-selected clusters in the eFEDS field \citep{Ramos-Ceja22} and the luminosity-mass-and-redshift relation of the eFEDS cluster sample with the HSC weak-lensing mass calibration \citep{Chiu22} within the measurement errors though the fitting functions are not exactly the same.

Finally, we discuss the interpretation of our results by referring to the baseline scaling relations. The self-similar model \citep{Kaiser86} predicts simple relations between X-ray properties of ICM and mass in the absence of baryonic physics such as AGN feedback and radiative cooling. Moreover, these baseline relations were derived without considering the fact that less massive clusters tend to be more concentrated and have higher characteristic densities in the hierarchical structure formation in a CDM universe \citep{NFW97}. On the other hand, \cite{Fujita19} constructed the new baseline luminosity-temperature and mass relations by considering the mass-concentration relationship. This is again the case when additional physics, such as feedback and radiative cooling, do not work.
They showed that the baseline relations should be shallower than the conventional self-similar model and follow $L\propto T^{1.6-1.8}$ and $L\propto M^{1.1-1.2}$. They also suggested that the $L-T$ relation of high-mass clusters should be close to the baseline model because the feedback from stars and AGN is less effective.

To carry out a comparison with their model, we derived the scaling relations without $E(z)$-correction in the same manner as in \cite{Fujita19}. From rows 5-8 of Table~\ref{tab:scaling}, the fitted slopes of $L-T$ and $L-M$ relations agree with the baseline models within the $1\sigma$ errors. Here, the baseline relations show little difference between merging and non-merging clusters \citep[Fig.~3 of][]{Fujita19}, while disturbed clusters dominate our sample. The nature of the fundamental plane can explain the above trend because the $L-T$ relation is close to the edge-on view of the fundamental plane and the clusters do not substantially deviate from the thin plane, even during a merging process \citep{Fujita18}.  Therefore, the small $L-T$ slope observed in the high-richness massive clusters agrees with the prediction of the revised baseline model by \cite{Fujita19}.
We plan to extend the analysis using the eROSITA and HSC surveys to test the shallow scaling relations.

\section{Summary}
Based on a joint analysis of the eROSITA/eFEDS and Subaru/HSC surveys, we studied the X-ray properties of 43 optically selected clusters with a high richness of $>40$ at $0.16<z<0.89$. Our major findings are as follows:

\begin{enumerate}
    \item We studied the cluster dynamical status by the X-ray-BCG offset and the gas concentration parameter and measured the morphology of member-galaxy distributions by the peak-finding method. As a result, we estimated the fraction of relaxed clusters to be 
    $2(<39)$\%, which is smaller than that of the X-ray-selected cluster samples.
    \item We performed the X-ray spectral analysis and weak-lensing mass measurement, deriving the scaling relations using the hierarchical Bayesian regression method. The luminosity-temperature relation is shallow; the slope is consistent with the predictions of the self-similar model and the baseline model incorporating the mass-concentration relation. The luminosity-mass relation also agrees with the two theoretical models cited in this work as well as what has been observed for the shear-selected clusters in the eFEDS field within the measurement errors.
\end{enumerate}

Our joint eROSITA and HSC study showed that the average X-ray properties of high-richness clusters are likely to be different from those found in the X-ray cluster samples. To improve the sample size, we plan to incorporate more than 900 objects with $15<N<40$ (Table~\ref{tab:sample}) and extend the analysis to the eROSITA all-sky survey (eRASS) data that overlap with the HSC footprint. These works enable us to study the mass-observable relations and the redshift evolution of the optical clusters. Furthermore, comparisons with the X-ray and shear-selected samples will improve our knowledge about the selection effect and cluster evolution.

\begin{acknowledgements}

This work is based on data from eROSITA, the soft X-ray instrument aboard SRG, a joint Russian-German science mission supported by the Russian Space Agency (Roskosmos), in the interests of the Russian Academy of Sciences represented by its Space Research Institute (IKI), and the Deutsches Zentrum f\"{u}r Luft- und Raumfahrt (DLR). The SRG spacecraft was built by Lavochkin Association (NPOL) and its subcontractors, and is operated by NPOL with support from the Max Planck Institute for Extraterrestrial Physics (MPE).

The development and construction of the eROSITA X-ray instrument was led by MPE, with contributions from the Dr. Karl Remeis Observatory Bamberg \& ECAP (FAU Erlangen-Nuernberg), the University of Hamburg Observatory, the Leibniz Institute for Astrophysics Potsdam (AIP), and the Institute for Astronomy and Astrophysics of the University of T\"{u}bingen, with the support of DLR and the Max Planck Society. The Argelander Institute for Astronomy of the University of Bonn and the Ludwig Maximilians Universit\"{a}t Munich also participated in the science preparation for eROSITA.

The eROSITA data shown here were processed using the eSASS/NRTA software system developed by the German eROSITA consortium.

The Hyper Suprime-Cam (HSC) collaboration includes the astronomical communities of Japan and Taiwan, and Princeton University. The HSC instrumentation and software were developed by the National Astronomical Observatory of Japan(NAOJ), the Kavli Institute for the Physics and Mathematics of the Universe (Kavli IPMU), the University of Tokyo, the High Energy Accelerator Research Organization (KEK), the Academia Sinica Institute for Astronomy and Astrophysics in Taiwan (ASIAA), and Princeton University. Funding was contributed by the FIRST program from Japanese Cabinet Office, the Ministry of Education, Culture, Sports, Science and Technology (MEXT), the Japan Society for the Promotion of Science (JSPS), Japan Science and Technology Agency (JST), the Toray Science Foundation, NAOJ, Kavli IPMU, KEK,ASIAA, and Princeton University.

We thank Y. Fujita for valuable discussions and the anonymous referee for comments.

This work was supported in part by the Fund for the Promotion of Joint International Research, JSPS KAKENHI Grant Number 16KK0101, 20K04027(NO), 20H05856, 20H00181, 19KK0076, 22H01260 (MO).
\end{acknowledgements}

\bibliographystyle{aa}
\bibliography{ref}

\end{document}